\documentclass[lettersize,journal]{IEEEtran}
\usepackage{amsmath,amsfonts}
\usepackage{algorithmic}
\usepackage{array}
\usepackage[caption=false,font=normalsize,labelfont=sf,textfont=sf]{subfig}
\usepackage{textcomp}
\usepackage{stfloats}
\usepackage{url}
\usepackage{verbatim}
\usepackage{graphicx}
\usepackage[colorlinks=true,linkcolor=red,citecolor=blue,urlcolor=blue,]{hyperref}
\usepackage{booktabs}
\usepackage{multirow} 
\usepackage[capitalize]{cleveref}
\newtheorem{lemma}{Lemma}
\newtheorem{theorem}{Theorem}
\newtheorem{remark}{Remark}
\newtheorem{corollary}{Corollary}
\newtheorem{proposition}{Proposition}

\def\BibTeX{{\rm B\kern-.05em{\sc i\kern-.025em b}\kern-.08em
		T\kern-.1667em\lower.7ex\hbox{E}\kern-.125emX}}
\usepackage{balance}
\begin{document}
	\title{ Performance Analysis of Fluid Antenna Multiple Access Assisted Wireless Powered Communication Network }
	\author{Xiao Lin, Yizhe Zhao,~\IEEEmembership{Member,~IEEE,} Halvin Yang, and Jie Hu,~\IEEEmembership{Senior Member,~IEEE}
		\vspace{-9mm}
		\thanks{Xiao Lin, Yizhe Zhao and Jie Hu are with the School of Information and Communication Engineering, University of Electronic Science and Technology of China, Chengdu 611731, China (e-mail: xiaolin@std.uestc.edu.cn; yzzhao@uestc.edu.cn; hujie@uestc.edu.cn).}
		\thanks{Halvin Yang is with the Wolfson School of Mechanical, Electrical and Manufacturing Engineering, Loughborough University, Loughborough, LE11 3TU, United Kingdom (e-mail: h.yang6@lboro.ac.uk).}
	}

	\maketitle
	
	\begin{abstract}
		This paper investigates a novel wireless powered communication network (WPCN) enabled by fluid antenna multiple access (FAMA). In the proposed system, a hybrid access point (HAP) equipped with multiple fixed-position antennas (FPA) delivers integrated data and energy transfer (IDET) services. Each low-power device is equipped with a single fluid antenna (FA) and utilizes the harvested energy to support its uplink communication. A block-correlation fading channel model is adopted to analyze the outage probabilities of downlink and uplink wireless data transfer (WDT) under various port selection strategies, including downlink signal-to-interference ratio-based port selection (DSPS), downlink energy-harvesting-power-based port selection (DEPS), uplink signal-to-noise ratio-based port selection (USPS), and uplink channel-based port selection (UCPS). To facilitate analysis, a step function approximation (SFA) method is proposed and then employed to obtain tractable expressions, including simplified integral forms. Moreover, lower bounds for the uplink WDT outage probability are derived. Numerical results validate the accuracy of the theoretical analysis and provide useful insights for system design.
	\end{abstract}
	
	\begin{IEEEkeywords}
		Fluid antenna multiple access (FAMA), wireless powered communication network (WPCN), port selection strategy, outage probability.
	\end{IEEEkeywords}

	\section{Introduction}
	\subsection{Background}
Recent advances in wireless mobile communications have led to the emergence of multiple progressive technologies, such as multiple-input multiple-output (MIMO), which enhance network capacity and efficiency. However, most of the connected devices are tiny and energy-limited, which have both communication and energy replenishment requirements. Since it is cost-prohibitive and impractical to replace the batteries of these devices manually in some challenging environments, wireless energy transfer (WET) has emerged as a promising technology to extend the battery lifespan of such devices. By coordinating with wireless data transfer (WDT), the receiver is able to perform energy harvesting (EH) and information decoding (ID) simultaneously, which yields the concept of integrated data and energy transfer (IDET) \cite{8421584,6760603}. Further, wireless powered communication network (WPCN) is studied as a typical paradigm for serving all the low-power devices, which is capable of gleaning radio frequency (RF) energy for powering their own uplink transmission. With the aid of the technique of WET or IDET, the problem of energy limitations of wireless devices  can be readily alleviated \cite{10534278,7081084}.
	
	However, due to the limitation of the energy harvesting circuit, WET has a higher requirement on the receive signal strength compared to WDT, which is posing a challenge on the efficiency of wireless signals' transmission. Although the MIMO technology has significantly boomed wireless mobile communications, it is not suitable for small low-power devices due to its higher power consumption and increased hardware complexity. To overcome this bottleneck, fluid antenna (FA) systems have emerged as a viable technology. Unlike traditional MIMO, which requires multiple fixed antennas separated by sufficient distances to achieve spatial diversity, FA is able to achieve comparable diversity within a tiny space by flexibly adjusting the antenna shape and position \cite{9264694}. This characteristic is particularly advantageous for IDET in WPCN due to its ability to enhance the WDT and WET.
	
	The FA system can also be applied in the multi-user scenario, which yields the concept of fluid antenna multiple access (FAMA) \cite{9650760,10066316}. With the aid of FAMA, it is able to mitigate the interference and enhance the signal-to-interference-plus-noise ratio (SINR) by selecting the optimal port where  the signal envelopes from the other users were in a deep fade. However, as for the multi-user FAMA-assisted WPCN system, the port selection strategies should be quite different, since the interference at the receiver plays a different role for WET and WDT, \textit{i.e.}, the interference may negatively impact the information decoding, but can boost the energy harvesting performance. In order to unveil the trade-off between WET and WDT, this paper will focus on the multi-user scenario and investigate the performance analysis of the FAMA-assisted WPCN.
	
	\subsection{Related Works}
	Due to the hardware limitations, one of the challenges in designing IDET systems is that ID and EH cannot be implemented through the same circuit module. Therefore, time switching (TS) and power splitting (PS) are two basic approaches to address this issue \cite{6489506}. In the TS approach, the received signal is periodically directed for ID and EH via a time switcher, while in the PS approach, the signal received at the receiver is separated into two parts, one for ID and the other for EH. IDET has been widely studied in various scenarios. For instance, the ergodic throughput in fading relay channels was  investigated with IDET in \cite{7572122}, an adaptive power allocation scheme for IDET-enabled full-duplex cooperative non-orthogonal multiple access (NOMA) networks using a time-switching protocol was investigated in \cite{9170590}. Also, IDET was studied in the MIMO interference channel \cite{6861455}. On the other hand, WPCN  has also attracted significant attention, {\em e.g.}, Budhiraja \textit{et al.} \cite{9496104} maximized the energy efficiency for wireless powered based D2D communication systems, and  Ju \textit{et al.} \cite{6678102} proposed a novel common-throughput maximization approach for WPCN. Moreover, some other technologies, {\em e.g.}, MIMO \cite{7843670}, unmanned aerial vehicle (UAV) \cite{10032196}, and reconfigurable intelligent surface (RIS) \cite{10330059}  were also studied in the WPCN in order to enhance both the WDT and WET performance.  Besides, relay selection has also been considered in the WPCN, since it may directly influence the WDT and WET performance by coordinating different wireless channels. For instance, in \cite{8269400},  the asymptotic distribution of the throughput of the $k$-th best link over  fading channels was derived using extreme value theory. In \cite{8880479}, the system outage probability and reliable throughput were derived based on the joint optimal selection of  the $k$-th best relay and the transmit antenna.
	
	On the other hand, in order to improve the efficiency of wireless signals' transmission, FA has been considered in various wireless communication scenarios. Specifically, New \textit{et al.} in \cite{10130117} approximated the outage probability and diversity gain of the FA system in closed-form expressions into the single-input and single-output (SISO) system, while the achievable performance of the FA system-assisted MIMO systems was further studied in \cite{10303274}. Moreover, FAMA systems were also studied in \cite{10078147,10436574} to improve the spectrum efficiency of the system without utilizing additional communication resources. For instance, Wong \textit{et al.} studied the benefits of the synergy between opportunistic scheduling and FAMA \cite{10078147}, and Xu \textit{et al.} investigated the outage performance of a downlink FAMA system having two users under a fully correlated channel model \cite{10436574}.
	
	The FA system has also been investigated by integrating the technology of IDET \cite{10622204,lin2024fluid,10622373,10506795,10615443,ghadi2024performancefasWP} in WPCN. These studies have provided significant insights into the potential of FA in WET scenarios. Specifically, the WDT as well as WET performance of the FA-assisted IDET system were evaluated theoretically by conceiving the TS approach \cite{10622204} and PS approach \cite{lin2024fluid,10622373}. Moreover, Zhang \textit{et al.} optimized the weighted energy harvesting power of energy receivers by satisfying the SINR constraints for each data receiver, which are all equipped with a single FA \cite{10506795}. Lai \textit{et al.} first derived the analytical outage probability of the one-dimensional (1D) FA-assisted wireless powered communication system \cite{10615443}, while Ghadi \textit{et al.} studied the performance of the 2D FA-assisted WPCN under the NOMA scheme \cite{ghadi2024performancefasWP}.
	\subsection{Motivations and Contributions}
	Despite the advantages in both WET and FA technologies, the study of FA in the WPCN remains uncharted territory, presenting several challenges: 1) The spatial correlation between FA ports complicates the derivation of outage probability, making performance analysis of the WPCN difficult; 2) The complex expression of outage probability hinders the provision of insights into future FA system design; 3) Since the port selection strategy in FA can be tailored to different requirements, its impact on downlink and uplink WDT performance in WPCN remains unclear; 4) The interference of the FAMA-assisted multi-user scenario has different impacts on the WDT and WET performance, which should be further investigated.
	
	In order to address these challenges, this paper aims to study the outage probability analysis of FAMA-assisted WPCN with a more practical wireless channel model by considering various port selection strategies. Specifically, in the considered system, a hybrid access point (HAP) with multiple traditional antennas communicates with multi-users equipped with a single FA. The PS approach is conceived at the receiver to simultaneously coordinate WDT and WET for realizing downlink IDET, while the receivers also use gleaned energy for powering their uplink transmission. Although the multi-user interference may degrade WDT performance, it can provide additional RF sources for EH to further boost the uplink transmission. Our main contributions of this paper are then summarized as follows: 
	\begin{itemize}
		\item We study an FAMA-assisted WPCN system, while the outage performance is analyzed theoretically under some typical port selection strategies, including: 1) downlink signal-interference-ratio based port selection (DSPS)  strategy; 2) downlink energy harvesting power based port selection (DEPS) strategy; 3) uplink channel based port selection (UCPS) strategy; and 4) uplink signal-to-noise ratio based port selection (USPS) strategy. Each strategy corresponds to different channel state information (CSI) requirements.
		\item  The closed-forms of the downlink WDT outage probabilities are obtained under the USPS, DEPS, and UCPS strategies, while an exact expression of that under the DSPS strategy is also formulated. For the uplink WDT outage probabilities, we derive exact expressions under the USPS, DEPS, and UCPS strategies, while an approximate closed-form solution for the DSPS strategy is obtained.
		\item A step function approximation (SFA) method is proposed to significantly reduce the computational complexity across all strategies, with its applicable conditions specified. Additionally, lower bounds on the performance under the DEPS, UCPS, and USPS strategies are provided to facilitate performance benchmarking and system design.
		\item Numerical results show that the DSPS strategy performs optimally in the downlink WDT, whereas the USPS strategy achieves the best performance among the proposed strategies in the uplink WDT. Furthermore, simulation results demonstrate the effectiveness of the proposed approximation method. The results also confirm that the FAMA system outperforms traditional fixed position antennas (FPA) system with selection combining (SC).
	\end{itemize}
	
The remainder of this paper is organized as follows. Section \ref{SystemModel} introduces the FAMA-assisted WPCN system, along with the block-correlation channel model, performance metrics, and port selection strategies. In Section \ref{Analysis}, we analyze the downlink and uplink WDT outage probabilities under different port selection strategies. Section \ref{umerical Results} validates the theoretical analysis through numerical simulations. Finally, Section \ref{Conclusion} concludes the paper.

	\section{System Model}\label{SystemModel}
	We consider a FAMA-assisted WPCN system consisting of an HAP and $M$ users, where the HAP is equipped with $M$ FPA and  each user is equipped with a single FA having $N$ ports\footnote{The FA structure can be a linear (1D) or planar (2D) FA based on radio frequency (RF) pixels.}. Each HAP antenna serves a corresponding user for WDT, \textit{i.e.}, the $m$-th HAP antenna communicates with user $m$. The timeline is divided into several periods, each lasting $T$. At the beginning of each period, the user activates one port based on a specific port selection strategy\footnote{Given that an RF-based pixel FA is considered, it is reasonable to assume that the energy consumed during port switching is negligible.}, and it remains unchanged throughout the period for the execution of downlink IDET and uplink WDT. Then, during the downlink IDET phase\footnote{This paper considers an RF pixel-based FA, thus, it is reasonably assumed that the time for port selection is negligible.} having the duration  of $t_1$, $0 < t_1 < T$, the HAP transmits dedicated RF signals towards all the users via their corresponding antennas, while each user applies the PS approach for realizing downlink IDET. Subsequently, during the uplink WDT phase having the duration of $t_2$ $\left( t_2 = T -t_1\right) $, each user uploads information to the HAP by using all the harvested energy via the same port. Note that since the HAP is equipped with traditional antennas, it is unable to coordinate multi-user interference via FAMA. Therefore, in the uplink phase, all the users upload their own information using orthogonal resources to avoid any interference at the HAP. Without loss of generality, the following  analysis concerns the typical user $m$ but the results hold for all the users of the proposed system.
	\begin{figure}[t]
		\centering
		\includegraphics[width=3in]{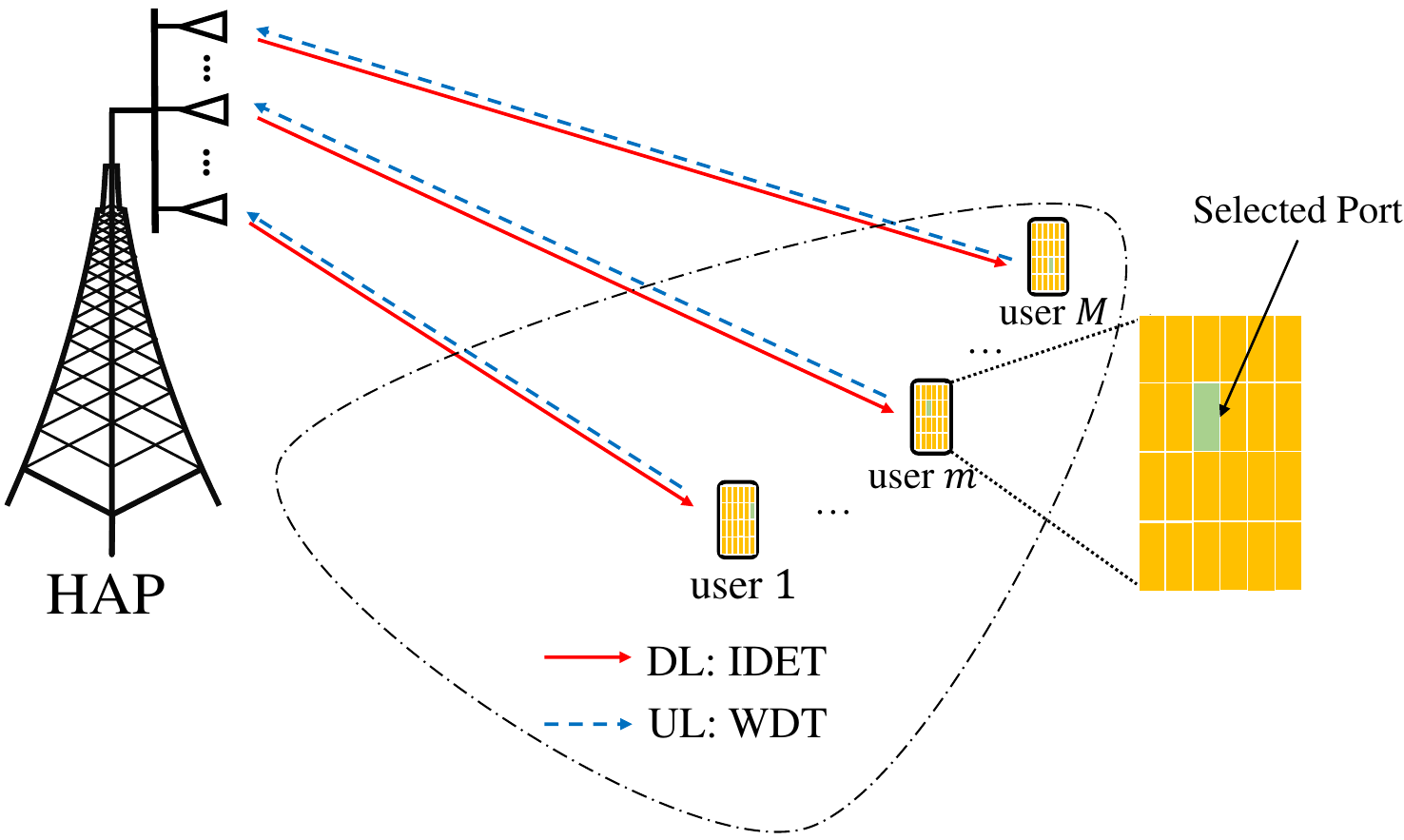}
		\caption{A system of FAMA-assisted WPCN.}
		\label{Model}
	\end{figure}
	\begin{figure}[t]
		\centering
		\includegraphics[width=2.5in]{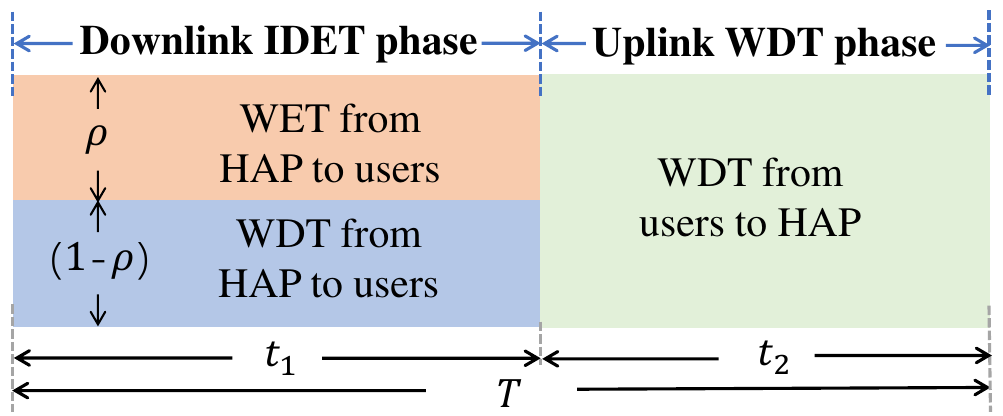}
		\caption{A transmission framework of the FAMA-assisted WPCN system.}
		\label{TransmissionBlock}
	\end{figure}
	\subsection{Signal Model}
	In the downlink IDET phase, user $m$ employs the PS approach to coordinate WDT and WET from the received signal. Specifically, the received signal at each UE is split by a power splitter which divides $\rho^{(m)} $ $(0 \leq \rho^{(m)} \leq 1)$ portion of the signal power for WDT, and the remaining $(1-\rho^{(m)})$ portion of power for WET. Therefore, the received WDT signal at the $n$-th port FA of user $m$ is
	\begin{multline}
		r_{\text{D},n}^{(m)}=s^{(m)}_{\text{D}}\sqrt{\frac{\rho^{(m)} P^{(m,m)}_{\text{D}}}{\Omega}}g_n^{(m,m)}+\\\sum_{\tilde{m}\neq m \atop\tilde{m}=1}^Ms^{(\tilde{m})}_{\text{D}}\sqrt{\frac{\rho^{(m)} P^{(\tilde{m},m)}_{\text{D}}}{\Omega}}g_n^{(\tilde{m},m)}+\omega_{\text{D},n}^{(m)},
	\end{multline}
	where $s^{(m)}_{\text{D}} \in \mathbb{C} $ is the downlink signal intended for user $m$ having the unit power, \textit{i.e.},  $\mathbb{E}[\vert s^{(m)}_{\text{D}}\vert^2]=1$, $P_D^{(\tilde{m},m)}$ is the transmit power of $\tilde{m}$-th HAP antenna and is assumed to be the same as $P_t$ for each transmit antenna without loss of generality, $g_n^{(\tilde{m},m)} \forall n\in \mathcal{N}$ is the channel coefficient from the $\tilde{m}$-th HAP antenna to the $n$-th port of user $m$'s FA, $w_{\text{D},n}^{\left(m\right)} \sim \mathcal{CN}(0,\sigma_{w}^{2})$ is the additive white Gaussian noise (AWGN) at the FA of user $m$, $\Omega$ is the path loss\footnote{Without loss of generality, it is assumed that the pathloss between the HAP and different users remains the same.}. Then, by applying the PS approach, the energy harvesting power (EHP) at the $n$-th port of user $m$,  accounting for the energy conversion efficiency conversion $\eta$ is given by
	\begin{equation}\label{EHP-multiuser}
		E_n^{(m)} = \eta\left( 1-\rho^{(m)}\right) \frac{P_tt_1}{\Omega}\sum_{\tilde{m}=1}^M\vert g_n^{(\tilde{m},m)} \vert^2.
	\end{equation}
	Accordingly, by neglecting the noise item, the received signal-to-interference ratio (SIR)\footnote{In this paper, we assume that the system is operating in an interference-limited scenario. Thus, the interference power is much greater than the noise power, and SIR serves as a suitable approximation for the SINR.} at the $n$-th port of user $m$ is given by
	\begin{equation}\label{SIR}
		\gamma^{(m)}_{\text{D},n}=\frac{\vert g_n^{\left(m,m\right)} \vert^2}{\sum_{\tilde{m}\neq m \atop\tilde{m}=1}^M\vert g_n^{(\tilde{m},m)} \vert^2}.
	\end{equation}
	
	During the uplink WDT phase, the received signal at the HAP from the $n$-th port of the typical user $m$ is expressed as
	\begin{equation}
		r_{\text{U},n}^{(m)} = s_{\text{U}}^{(m)} \sqrt{\frac{P_n^{(m)}}{\Omega}}h_n^{\left(m\right)}+w_{\text{U}}^{(m)},
	\end{equation}
	where $P_n^{(m)} = \frac{E_n^{(m)}}{t_2}$ is the transmission power available for the uplink WDT phase. $ h_n^{\left(m\right)}$ is the uplink multi-path fading coefficient from the $n$-th fluid antenna port of the user $m$ to the corresponding antenna of the HAP. $s_{\text{U}}^{(m)}$ is the uplink information signal with $\mathbb{E}\left[\vert s_{\text{U}}^{(m)}\vert^2\right]=1$, and $w_{\text{U}}^{(m)} \sim \mathcal{CN}\left(0,\sigma_w^2\right)$ is the AWGN at the HAP antenna. Since there is no additional interference during the uplink transmission, the signal-to-noise ratio (SNR) at the HAP for receiving the user $m$'s uplink signal is given as
	\begin{equation}\label{SNR}
		\gamma^{(m)}_{\text{U},n} = \frac{\vert h_n^{(m)}\vert^2}{t_2\sigma_{w}^2\Omega}E_n^{(m)}.
	\end{equation}
	\subsection{Wireless Channel Model}
	The spatial separation among the ports of FA yields a difference in the phases of arriving
	paths, thus inducing correlation between the channels following Jake’s model  between the $n_1$-th port and the $n_2$-th port, which is modeled as \cite{10103838}
	\begin{equation}\label{trueEigMatrix}
		\left(\boldsymbol{\Sigma}\right)_{n_1,n_2} = J_0 \left( \frac{2\pi (n_1-n_2)W}{N-1} \right), n_1,n_2 \in \mathcal{N},
	\end{equation}
	where $J_0(\cdot)$ is the zero-order Bessel function of the first kind and $W$ is the antenna size of FA. Although the model can accurately characterize the correlation, it makes the analysis challenging to conduct. To address this issue, \cite{10623405} proposed the block-correlation model to simplify the representation of \eqref{trueEigMatrix}. In this model, different blocks are defined to be independent, while the spatial correlation within each block remains constant. The spatial correlation over the ports is characterized as
	\begin{equation}
		\widehat{\boldsymbol{\Sigma}}\in\mathbb{R}^{N\times N}=\begin{pmatrix}
			\mathbf{A}_1&\mathbf{0}&\ldots&\mathbf{0}\\\mathbf{0}&\mathbf{A}_2&\ldots&\mathbf{0}\\\vdots&&\ddots&\vdots\\\mathbf{0}&\mathbf{0}&\mathbf{0}&\mathbf{A}_B
		\end{pmatrix},
	\end{equation}
	where  each submatrix $\mathbf{A}_b$ is a constant correlation matrix having the size of $L_b$ and correlation of $\mu^2$, which is given by
	\begin{equation}
		\mathbf{A}_b\in\mathbb{R}^{L_b\times L_b}=\begin{pmatrix}1&\mu^2&\ldots&\mu^2\\\mu^2&1&\ldots&\mu^2\\\vdots&&\ddots&\vdots\\\mu^2&\ldots&\mu^2&1\end{pmatrix},\quad b=1,\ldots,B.
	\end{equation}
	where $\mu^2 $ is a constant close to $1$, $\sum_{b=1}^BL_b=N$. The block sizes $L_b$ and block number $B$ are chosen based on spectral analysis of
	the true correlation matrix in \eqref{trueEigMatrix}, and the detailed information can be found in \cite{10623405}. Thus, the wireless channel between the $m$-th antenna at HAP and the user $m$  in the  downlink IDET phase can be re-expressed as\footnote{In this paper, we assume that the system is operating in a rich scattering environment.}
	\begin{multline}\label{DLchannel}
		g_{n,b(n)}^{(m,m)} = \sqrt{1-\mu^2}x_{n,b(n)}^{(m,m,\text{D})} + \mu x_{b(n)}^{(m,m,\text{D})} \\+ j\left(\sqrt{1-\mu^2}y_{n,b(n)}^{(m,m,\text{D})}+ \mu y_{b(n)}^{(m,m,\text{D})}\right),
	\end{multline}
	and the wireless channel between the user $m$ and the HAP in the uplink WDT phase can be re-expressed as
	\begin{multline}\label{ULchannel}
		h_{n,b(n)}^{(m)} = \sqrt{1-\mu^2}x_{n,b(n)}^{(m,\text{U})} + \mu x_{b(n)}^{(m,\text{U})} \\+ j\left(\sqrt{1-\mu^2}y_{n,b(n)}^{(m,\text{U})}+ \mu y_{b(n)}^{(m,\text{U})}\right),
	\end{multline}
	where $x_{n,b(n)}^{s}, x_{b(n)}^{s}, y_{n,b(n)}^{s}, y_{b(n)}^{s}, s\in\{(m,m,\text{D}),(m,\text{U})\}$ are all independent Gaussian variables with zero mean and variance of $1$. $b(n)$ is the block index, \textit{i.e.}, $b(n) = 1$ for $n=1,\ldots,L_1$, $b(n) = 2$ for $n=L_1+1,\ldots,L_1+L_2$, and so on \cite{10623405}. For simplicity of notation, we use $n$ to represent $n,b(n)$ in the following, \textit{e.g.,} $x_{n}^{(s)}$ represents $x_{n,b(n)}^{(s)}$.  In this paper, the outage performance of the FAMA-assisted WPCN system will be analyzed based on the block-correlation model described above.
	\subsection{Performance Evaluation}
	The performance of the proposed FAMA-assisted WPCN is analyzed in terms of the WDT outage probability for both the downlink and uplink phases.  A downlink WDT outage occurs when the received SIR at a typical user $m$ falls below a certain threshold, which is defined as
	\begin{equation}
		P\left(\gamma^{\text{D}}_{\text{th}}\right)=\text{Pr}\left(\gamma_{\text{D},n^{\ast}}^{(m)}=\frac{\vert g_{n^{\ast}}^{(m,m)} \vert^2}{\sum_{\tilde{m}\neq m \atop\tilde{m}=1}^M\vert g_{n^{\ast}}^{(\tilde{m},m)} \vert^2}<\gamma^{\text{D}}_{\text{th}}\right),		
	\end{equation}
	where $n^{\ast}$ is the selected port determined by the specific port selection strategies and $\gamma^{\text{D}}_{\text{th}}$ is the minimum required SIR threshold for WDT in the downlink IDET phase. Similarly, an uplink WDT outage occurs when the received SNR at the HAP is below a certain threshold, which is defined as
	\begin{equation}
		P\left(\gamma^{\text{U}}_{\text{th}}\right)=\text{Pr}\left(\gamma^{(m)}_{\text{U},n^{\ast}}=\frac{\vert h_{n^{\ast}}^{(m)}\vert^2}{t_2\sigma_{w}^2\Omega}E_{n^{\ast}}^{(m)}<\gamma^{\text{U}}_{\text{th}}\right),		
	\end{equation}
	where $\gamma^{\text{U}}_{\text{th}}$ is the minimum required SNR threshold in the uplink WDT.
	\subsection{Port Selection Strategies}
	In each transmission period, the process consists of two phases: the downlink IDET and the uplink WDT. To enhance system performance, it is essential to adopt appropriate port selection strategies that optimize either the downlink or uplink WDT performance. To optimize the downlink WDT performance, a straightforward method is to select the port that maximizes the received SIR, resulting in the DSPS strategy. Conversely, to improve the uplink WDT outage performance, one can select the port that maximizes the received SNR at the HAP. This leads to the USPS strategy. However, it is important to note that the received SNR at the HAP depends on both the downlink harvested energy during the IDET phase and the uplink channel quality. Therefore, the use of USPS requires knowledge of both downlink and uplink CSI across all ports. This requirement inevitably increases the system complexity and energy consumption, which is not desirable for low-power devices.
		To mitigate this issue, we consider two simplified variants of the USPS strategy, namely the DEPS strategy and the UCPS strategy. These two strategies only rely on single-side CSI (\textit{i.e.,} either downlink or uplink), thus significantly reducing the port selection complexity at the expense of some performance loss compared to USPS.

	\section{Outage Probability Analysis}\label{Analysis}
	In the following subsections, we analyze the downlink and uplink WDT outage probabilities under the DSPS, DEPS, UCPS, and USPS strategies, respectively. To facilitate the analysis and gain more insights, we first introduce the SFA method as follows.
	\subsection{SFA method}\label{SFAmethod}
We first recall the definition of the generalized Marcum-$Q$ function, which plays a central role in the statistical characterization of wireless fading channels. The generalized Marcum-$Q$ function of order $p$ is defined as \cite[Eq. (4.60)]{simon2004digital}
\begin{equation}
	Q_p(a, b) = \int_b^\infty x \left( \frac{x}{a} \right)^{p-1} \exp\left( -\frac{x^2 + a^2}{2} \right) I_{p-1}(a x) \, dx,
\end{equation}
where $a, b > 0$, and $I_{p-1}(\cdot)$ denotes the $(p-1)$-th order modified Bessel function of the first kind.  However, since it typically appears inside integrals, further analysis becomes highly intractable. To enable analytical tractability, we introduce the following approximation.
		\begin{lemma}\label{lemma:semiAppro}
			The expression $\left[1 - Q_p(a, b)\right]^L$ can be approximated by a step function with a threshold $\delta(b, L)$, such that
			\begin{equation}\label{eq:semiAppro}
				\left[1 - Q_p(a, b)\right]^L \approx
				\begin{cases}
					0, & a > \delta(b, L) \\
					1, & a < \delta(b, L)
				\end{cases}
			\end{equation}
			where the threshold $\delta(b, L)$ is given by
			\begin{equation}\label{lemma1:threshold}
				\delta(b, L) = b + \frac{\frac{L - 1}{\sqrt{2\pi}}b + p - \frac{1}{2}}{\frac{(L - 1)\left(p - \frac{1}{2}\right)}{\sqrt{2\pi}} + \frac{p - \frac{1}{2}}{b} - b}.
			\end{equation}
		\end{lemma}
		
		\begin{IEEEproof}
			Please refer to Appendix~\ref{prooflemma:SFA}.
		\end{IEEEproof}

		\begin{lemma}\label{lemma:semiApproL=1}
			For the special case of \( L = 1 \), the function \( 1 - Q_p(a, b) \) can be approximated by a step function with threshold \( \delta(b) \), given by
			\begin{equation}\label{eq:semiApproL=1}
				1 - Q_p(a, b) \approx
				\begin{cases}
					0, & a > \delta(b) \\
					1, & a < \delta(b)
				\end{cases}
			\end{equation}
			where the threshold is
			\begin{equation}\label{eq:threshold_L=1}
				\delta(b) = \frac{b + \sqrt{b^2 + 4p - 2}}{2}.
			\end{equation}
		\end{lemma}
		
		\begin{IEEEproof}
			This lemma can be viewed as a special case of Lemma~\ref{lemma:semiAppro} with \( L = 1 \). For this case, the expression simplifies to \( 1 - Q_p(a, b) \). To determine the step-function threshold, we consider the point where the curvature of \( Q_p(a, b) \) changes most rapidly by solving
			\begin{equation}
				a^{\ast} = \arg\left\{ \frac{\partial^2 Q_p(a, b)}{\partial a^2} = 0 \right\}.
			\end{equation}
			The first derivative of \( Q_p(a, b) \) with respect to \( a \) is given by
			\begin{equation}\label{eq:first_derivative_l1}
				\frac{\partial Q_p(a,b)}{\partial a} = \frac{b^p}{a^{p-1}} e^{-\frac{a^2 + b^2}{2}} I_p(ab).
			\end{equation}
			For large $ab$, the asymptotic form of $I_p(ab)$ in \eqref{AsymptoticI} can be used. The resulting second derivative is approximated as
			\begin{equation}\label{SecondDerieveL=1}
				\frac{\partial^2 Q_p(a, b)}{\partial a^2} \approx
				\frac{b^{p - \frac{1}{2}}}{a^{p - \frac{1}{2}}} e^{-\frac{(a - b)^2}{2}} + \frac{p - \frac{1}{2}}{a} + a - b.
			\end{equation}
			
			By applying a first-order Taylor expansion around $a = b$, and setting the expression to zero, we obtain the analytical threshold as
			\begin{equation}
				\delta(b) = \frac{b + \sqrt{b^2 + 4p - 2}}{2}.
			\end{equation}
			This concludes the proof.
	\end{IEEEproof}

		\begin{figure*}[htbp]
			\begin{minipage}[t]{0.3\textwidth}
				\centering
				\includegraphics[width=\textwidth]{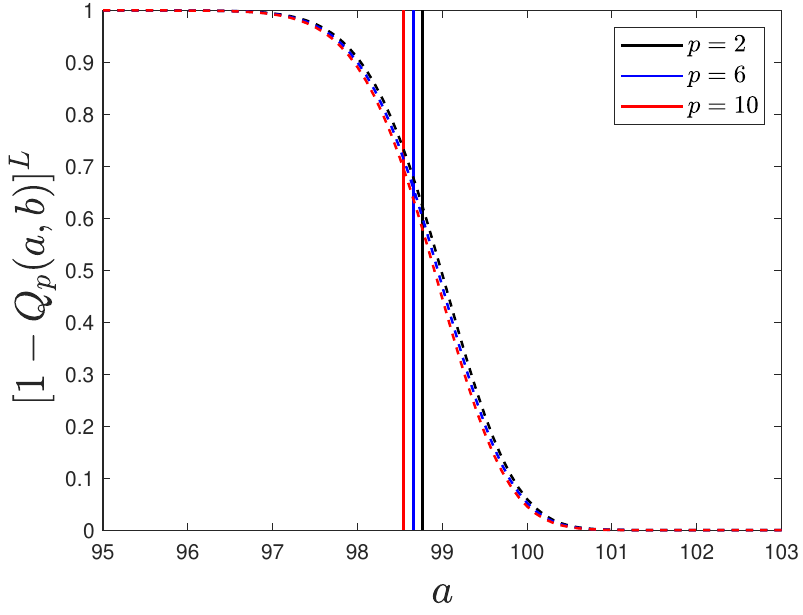}  
				\caption{Comparison of the threshold for different values of  different values with $p$ with $b = 100$ and $L=4$.}  
				\label{lemma1_approxp}
			\end{minipage} \hfill
			\begin{minipage}[t]{0.3\textwidth}
				\centering
				\includegraphics[width=\textwidth]{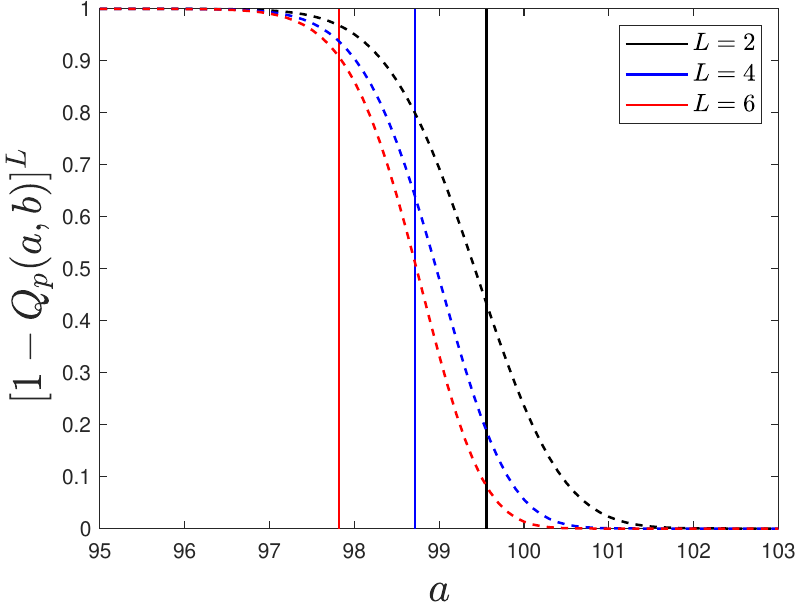}  
				\caption{Comparison of the threshold for different values of  different values with $L$ with $b = 100$ and $p=4$.}  
				\label{lemma1_approxL}
			\end{minipage} \hfill
			\begin{minipage}[t]{0.3\textwidth}
				\centering
				\includegraphics[width=\textwidth]{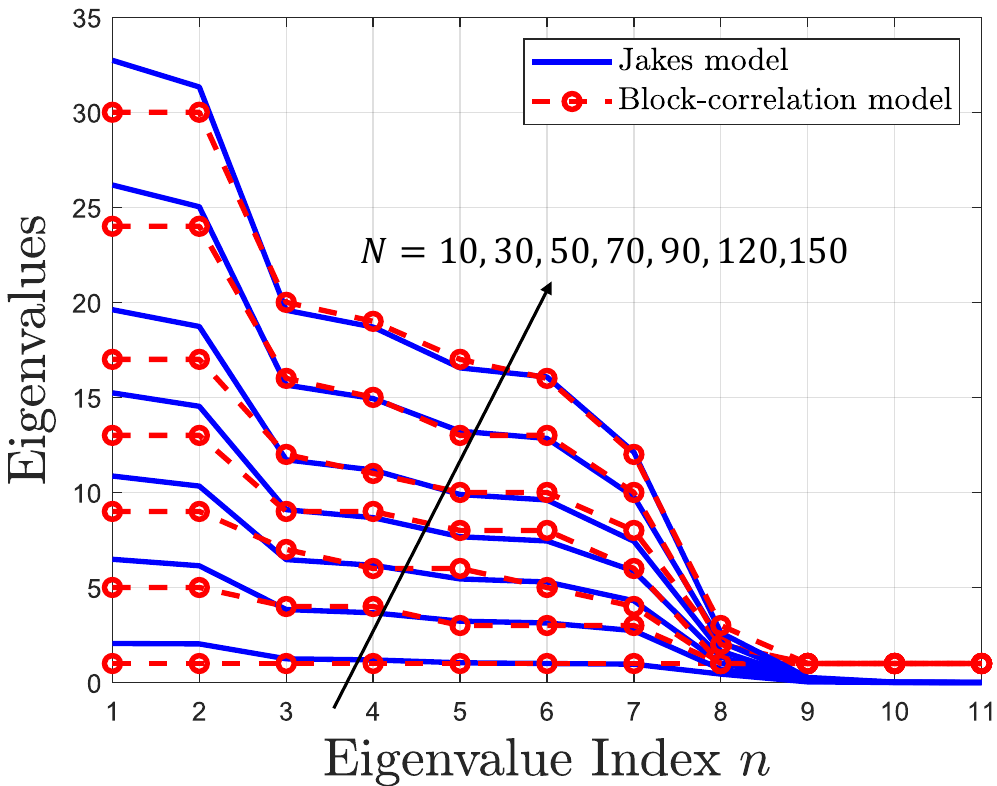}  
				\caption{Comparison of the eigenvalues for different values of  $N$ with $W=3$.}  
				\label{lemma1_approxEig}
			\end{minipage}
		\end{figure*}

		Before the proposed approximation is applied, its applicable range must be clearly defined. The approximation is valid when both $a$ and $b$ are relatively large. This condition arises from the use of the asymptotic expression of the modified Bessel function of the first kind in \eqref{AsymptoticI}. Fig.~\ref{lemma1_approxp} illustrates the impact of the parameter $p$ on the approximation accuracy. As $p$ increases, the accuracy of the approximation tends to deteriorate. This degradation is mainly due to the reduced precision of the asymptotic form for large values of $p$ \cite[Eq. (10.40.1)]{olver2010nist}.

		The impact of the parameter $L$ on the approximation accuracy is illustrated in Fig.~\ref{lemma1_approxL}. As $L$ increases, the approximation accuracy deteriorates due to the increasing nonlinearity of \eqref{linearFunction}. Under such condition, the valid range of the first-order Taylor expansion becomes narrower, and higher-order terms begin to play a significant role, resulting in larger approximation errors.

	To further assess the practical applicability, Fig.~\ref{lemma1_approxEig} compares the eigenvalues obtained from the block-correlation model, denoted by $L_b$, and those from the Jakes correlation model.  As shown, the block-correlation model maintains good accuracy as $N$ increases from 10 to 150. However, with a fixed antenna size, increasing $N$ degrades the SFA accuracy. For example, when $N = 60$, only about 25\% of the eigenvalues are above 10, whereas for $N = 150$, approximately 87.5\% exceed 10. This indicates that a larger $N$ results in more dominant eigenvalues in $L_b$ under fixed $W$, thereby reducing the accuracy of SFA.
		
		Moreover, the approximation performance of the block-correlation channel model degrades under very low antenna density (e.g., $N/W = 2$), where the eigenvalues tend to be small. Since the block-correlation model is primarily accurate for eigenvalues greater than or equal to 1, the approximation becomes unreliable in such cases, resulting in significantly reduced accuracy. Based on these observations, the proposed approximation method and the employed block-correlation channel model are more suitable for FA configurations with moderate antenna density. For example, when the antenna density falls within the range of $N/W= 5\sim20$, the approximation based on the block-correlation channel model achieves good accuracy.
	
	\subsection{Outage Probabilities under the DSPS Strategy}
	In the DSPS strategy, the user selects the port having the highest SIR, without considering the uplink channel.  When the user $m$ applies the DSPS strategy, the optimal antenna port $n^{\ast}$ is selected to maximize the $\gamma_{\text{D},n}^{(m)}$ as
	\begin{equation}\label{WPS-multi-user}
		n^{\ast}=\arg\max_{n} \frac{ \vert g_n^{(m,m)} \vert^2}{\sum_{\tilde{m}\neq m}^M \vert g_n^{(\tilde{m},m)} \vert^2}=\arg\max_{n}\frac{X_n}{Y_n}.
	\end{equation}
	where $X_n$ and $Y_n$ are defined as
	\begin{multline}
		X_n =\left(x_n^{(m,m,\text{D})}+\frac{\mu x_{b(n)}^{(m,m,\text{D})}}{\sqrt{1-\mu^2}} \right)^2\\+\left(y_n^{(m,m,\text{D})}+\frac{\mu y_{b(n)}^{(m,m,\text{D})}}{\sqrt{1-\mu^2}} \right)^2,
	\end{multline}
	\begin{multline}
		Y_n =\sum_{\tilde{m}\neq m \atop\tilde{m}=1}^M\left(x_n^{(\tilde{m},m,\text{D})}+\frac{\mu x_{b(n)}^{(\tilde{m},m,\text{D})}}{\sqrt{1-\mu^2}} \right)^2+\\
		\left(y_n^{(\tilde{m},m,\text{D})}+\frac{\mu y_{b(n)}^{(\tilde{m},m,\text{D})}}{\sqrt{1-\mu^2}} \right)^2.
	\end{multline}
	Thus, the downlink WDT outage probability is expressed as
	\begin{equation}
		P_{\text{DSPS}}^{\text{D}}\left(\gamma_{\text{th}}^{\text{D}} \right) =\text{Pr}\left(\arg\max_n \frac{X_n}{Y_n}<\gamma_{\text{th}}^{\text{D}}\right).
	\end{equation}
	\begin{theorem}\label{th:SIRWPSmultipleusers}
		By applying the DSPS
		strategy, the downlink WDT outage probability is formulated as
		\begin{multline}\label{eq:SIRWPSmultipleusers}
			P_{\text{DSPS}}^{\text{D}}\left(\gamma_{\text{th}}^{\text{D}} \right)=\prod_{b=1}^B\int_{0}^{\infty}\int_{0}^{\infty}\frac{\widetilde{r}_{b}^{M-2}e^{-r_{b}-\widetilde{r}_{b}}}{\Gamma(M-1)}\times\\\left[G\left(\gamma_{\text{th}}^{\text{D}};2r_{b},2\widetilde{r}_{b}\right)\right]^{L_b} dr_{b} d\widetilde{r}_{b},
		\end{multline}
	where $r_{b}$ and $\widetilde{r}_{b}$ are the non-central parameters, $G\left(\gamma; r_{b}, \widetilde{r}_{b} \right)$ is given in \eqref{eq:G}, and $\Gamma\left(\cdot\right)$ denotes the Gamma function.
		\begin{figure*}[t]
			\begin{multline}\label{eq:G}
				G(\gamma^{\text{D}}_{\text{th}};r_b,\widetilde{r}_b)=Q_{M-1}\left(\sqrt{\frac{\mu^2\gamma^{\text{D}}_{\text{th}}\widetilde{r}_b}{(1-\mu^2)(\gamma^{\text{D}}_{\text{th}}+1)}},\sqrt{\frac{\mu^2r_b}{(1-\mu^2)(\gamma^{\text{D}}_{\text{th}}+1)}}\right)-\left(\frac{1}{\gamma^{\text{D}}_{\text{th}}+1}\right)^{M-1}\exp\left(-\frac{\mu^2}{2(1-\mu^2)}\frac{\gamma^{\text{D}}_{\text{th}}\widetilde{r}_b+r_b}{\gamma^{\text{D}}_{\text{th}}+1}\right)\\
				\times\sum_{l=0}^{M-2}\sum_{j=0}^{M-l-2}\frac{(M-(j+l)-1)_j}{j!}\left(\frac{r_b}{\widetilde{r}_b}\right)^{\frac{j+l}2}(\gamma^{\text{D}}_{\text{th}}+1)^l\left( \gamma^{\text{D}}_{\text{th}}\right) ^{\frac{j-l}2}I_{j+l}\left(\frac{\mu^2\sqrt{\gamma^{\text{D}}_{\text{th}} r_b\widetilde{r}_b}}{(1-\mu^2)(\gamma^{\text{D}}_{\text{th}}+1)}\right)
			\end{multline}
			\hrulefill
		\end{figure*}
	\end{theorem}
	\begin{IEEEproof}
		The proof can be found in \cite[Appendix B]{10623405}.
	\end{IEEEproof}

	As observed in \eqref{eq:SIRWPSmultipleusers}, the expression is mathematically intractable due to the complexity of the involved integrations and summations, making it difficult to obtain any insightful results. With the SFA in hand, we are now prepared to derive an approximate expression for the downlink WDT outage probability under the DSPS strategy.

		\begin{corollary}\label{corollary:SIRmaxSIR}
			As $\mu^2 \rightarrow 1$, the downlink WDT outage probability under the DSPS strategy can be approximated as
			\begin{multline}\label{eqSFA:SIRmaxSIR}
				P_{\text{DSPS}}^{\text{D}}\left(\gamma_{\text{th}}^{\text{D}}\right) = \prod_{b=1}^{B} \int_{0}^{\infty} \int_{0}^{\infty} \frac{\widetilde{r}_{b}^{M-2} e^{-r_{b} - \widetilde{r}_{b}}}{\Gamma(M-1)} \\
				\times \left[ Q_{M-1} \left( \sqrt{\frac{2\mu^{2} \widetilde{r}_{b}}{1 - \mu^{2}}}, \delta_{\text{DSPS}}(r_b) \right) \right]^{L_b} \, dr_{b} \, d\widetilde{r}_{b},
			\end{multline}
			where the threshold function $\delta_{\text{DSPS}}(r_b)$ is defined as
			\begin{equation}\label{deltaDSPS}
				\delta_{\text{DSPS}}(r_b) = \sqrt{ \frac{\mu^{2} r_b}{2(1 - \mu^{2}) \gamma_{\text{th}}^{\text{D}}} } + \sqrt{ \frac{\mu^{2} r_b}{2(1 - \mu^{2}) \gamma_{\text{th}}^{\text{D}}} + \frac{1}{2 \gamma_{\text{th}}^{\text{D}}} }.
			\end{equation}
		\end{corollary}
		\begin{IEEEproof}
			Please refer to Appendix \ref{prooflemma:SIRmaxSIR}.
	\end{IEEEproof}
	
		\begin{corollary}\label{corollary:SIRmaxSIRSFAtwice}
			As $\mu^2 \rightarrow 1$, the downlink WDT outage probability under the DSPS strategy can be further approximated as
			\begin{equation}\label{eqSFAtwice:SIRmaxSIRtwice}
				P_{\text{DSPS}}^{\text{D}}\left(\gamma_{\text{th}}^{\text{D}}\right) \approx \prod_{b=1}^{B} \int_{0}^{\infty} e^{-r_b} \frac{\Gamma\left(M-1,\tilde{\delta}_{\text{DSPS}}\left(r_b,L_b\right)\right)}{\Gamma(M-1)} dr_{b},
			\end{equation}
			where $\Gamma\left(\cdot,\cdot\right)$ is the upper incomplete gamma function, and the threshold is
			\begin{equation}\label{DSPSthresholdtwice}
				\tilde{\delta}_{\text{DSPS}}\left(r_b,L_b\right) = \frac{1-\mu^2}{2\mu^2} \left[ \delta_{\text{DSPS}}(r_b) + \frac{\frac{L_b - 1}{\sqrt{2\pi}} + \frac{M - \frac{3}{2}}{\delta_{\text{DSPS}}(r_b)}}{\frac{L_b - 1}{\sqrt{2\pi}} \frac{M - \frac{3}{2}}{\delta_{\text{DSPS}}(r_b)} + 1} \right]^2.
			\end{equation}
			and $\delta_{\text{DSPS}}(r_b)$ is given in \eqref{deltaDSPS}.
		\end{corollary}
		\begin{IEEEproof}
			The expression in \eqref{eqSFA:SIRmaxSIR} contains the term $Q_p(a,b)^{L_b}$, which can be approximated by performing a first-order Taylor expansion around \( a = b \) \cite[Eq. (61)]{10623405}. Specifically, this yields
			\begin{equation}
				a \approx b + \frac{\frac{L_b - 1}{\sqrt{2\pi}} - \frac{p - \frac{1}{2}}{b}}{\frac{L_b - 1}{\sqrt{2\pi}} \cdot \frac{p - \frac{1}{2}}{b} + 1}.
			\end{equation}
			By substituting \( b = \delta_{\text{DSPS}}(r_b) \) and \( p = M - 1 \), and applying algebraic simplifications, we obtain the modified threshold in \eqref{DSPSthresholdtwice}. Substituting this result into the expression of \eqref{eqSFA:SIRmaxSIR}, we rewrite the outage probability as
			\begin{multline}
				P_{\text{DSPS}}^{\text{D}}\left(\gamma_{\text{th}}^{\text{D}}\right) = \prod_{b=1}^{B} \int_{0}^{\infty} \int_{\tilde{\delta}_{\text{DSPS}}\left(r_b,L_b\right)}^{\infty} \frac{\widetilde{r}_{b}^{M-2} e^{-r_{b} - \widetilde{r}_{b}}}{\Gamma(M-1)} \, d\widetilde{r}_{b} \, dr_{b}.
			\end{multline}
			Finally, by using the standard integral identity \cite[Eq. (3.351.2)]{gradshteyn2014table}, the inner integral simplifies to the upper incomplete gamma function, thereby leading to the single-integral form approximation in \eqref{eqSFAtwice:SIRmaxSIRtwice}. This completes the proof.
	\end{IEEEproof}
	\begin{remark}
		As $\mu^2 \rightarrow 1$, both parameters $a$ and $b$ in the generalized Marcum-$Q$ function $Q_p(a, b)$ tend to be large. In this regime, and under a high SIR threshold $\gamma_{\text{th}}^{\text{D}}$, the  threshold $\delta_{\text{DSPS}}(r_b)$ defined in \eqref{deltaDSPS} can be approximated as $\delta_{\text{DSPS}}(r_b) \approx \sqrt{ \frac{2\mu^2 r_b}{(1 - \mu^2)\gamma_{\text{th}}^{\text{D}}} }.$
			Since $\delta_{\text{DSPS}}(r_b)$ is asymptotically large in this case, the expression for $\tilde{\delta}_{\text{DSPS}}(r_b, L_b)$ in \eqref{DSPSthresholdtwice} simplifies to
			$	\tilde{\delta}_{\text{DSPS}}(r_b, L_b) \approx  \frac{1 - \mu^2}{2\mu^2} \left[ \delta_{\text{DSPS}}(r_b) + \frac{L_b - 1}{\sqrt{2\pi}} \right]^2\approx \frac{1 - \mu^2}{2\mu^2} \delta_{\text{DSPS}}^2(r_b) = \frac{r_b}{\gamma_{\text{th}}^{\text{D}}}$. In this case, a more simplified expression can be derived.
	\end{remark}

	Subsequently, we analyze the uplink WDT outage probability under the DSPS strategy, which can be expressed as
	\begin{equation}\label{firstSNRWPS}
		\begin{aligned}
			P_{\text{DSPS}}^{\text{U}}\left(\gamma_{\text{th}}^{\text{U}} \right) &=\text{Pr}\left(\gamma_{\text{U},n^{\ast}}^{(m)}<\gamma_{\text{th}}^{\text{U}} \vert n^{\ast}=\arg\max_{n}\gamma_{\text{D},n}^{\left(m\right)}\right)\\
			&=\text{Pr}\left(\beta_{n^{\ast}}\alpha_{n^{\ast}}<\widetilde{\gamma}\vert n^{\ast}=\arg\max_{n}\frac{X_n}{Y_n}\right),
		\end{aligned}		
	\end{equation}
	where  $\widetilde{\gamma}=\frac{\gamma^{\text{U}}_{\text{th}} t_2\sigma_w^2\Omega^2}{\eta(1-\rho^{(m)}) P_tt_1\left(1-\mu^2\right)^2}$, $\alpha_n =X_n+Y_n$, and the variable $\beta_n$ is defined as
	\begin{equation}
		\beta_n =\left(x_n^{(m,\text{U})}+\frac{\mu x_{b(n)}^{(m,\text{U})}}{\sqrt{1-\mu^2}} \right)^2+\left(y_n^{\left(m,\text{U}\right)}+\frac{\mu y_{b(n)}^{\left(m,\text{U}\right)}}{\sqrt{1-\mu^2}} \right)^2.
	\end{equation}
	
	In order to evaluate the uplink WDT outage probability, we first present the following lemma.
	\begin{lemma}\label{lemma:XYindependent}
		When $\mu^2=1$, the variables $\frac{X_n}{Y_n}$ and $X_n+Y_n$ are  independent.
	\end{lemma}
	\begin{IEEEproof}
		The proof can be found in \cite[Appendix C]{lin2024fluid}.
	\end{IEEEproof}
	
	With the aid of Lemma \ref{lemma:XYindependent}, we can derive an approximate expression for the uplink WDT outage probability, as presented in the following theorem.
	\begin{theorem}\label{th:SNRWPSmultipleusers}
		The uplink WDT outage probability  under the DSPS strategy can be approximated as 
		\begin{equation}\label{eq:SNRWPSmultipleusers}
			P_{\text{DSPS}}^{\text{U}} \left(\gamma_{\text{th}}^{\text{U}} \right)      \approx1-\frac{\left[\widetilde{\gamma}\left(1-\mu^2\right)\right]^{\frac{M}{2}}}{2^{M-1}\Gamma(M)}K_M\left(\sqrt{\widetilde{\gamma}\left(1-\mu^2\right)}\right),
		\end{equation}
		where $K_p\left(\cdot\right)$ is the $p$-th modified Bessel function of the second kind.
	\end{theorem}
	\begin{IEEEproof}
		Please refer to Appendix \ref{proofth:SNRmaxSIR}.
	\end{IEEEproof}
	\subsection{Outage Probabilities under the DEPS Strategy}
	In this subsection, we will analyze the outage performance under the DEPS strategy. In the DEPS strategy, the user adjusts the port based on the EHP at user $m$. In other words, the DEPS strategy selects the port that harvests the maximum amount of energy. If the user applies the DEPS strategy, the optimal antenna port $n^{\ast}$ is selected by following
	\begin{equation}
		n^{\ast} = \arg\max_nE_n^{(m)} = \arg\max_n \sum_{\tilde{m}=1}^M\left\vert g_n^{(\tilde{m},m)} \right\vert^2.
	\end{equation}
	Then, our first objective is to analyze the downlink WDT outage probability, which is expressed as
	\begin{equation}\label{eq:originalDEPS}
		\begin{aligned}
			P_{\text{DEPS}}^{\text{D}}\left(\gamma^{\text{D}}_{\text{th}}\right)&=\text{Pr}\left(\gamma_{\text{D},n^{\ast}}^{(m)}<\gamma^{\text{D}}_{\text{th}}| n^{\ast}=\arg\max_n  E_n^{(m)} \right),\\
			&=\text{Pr}\left(\frac{X_{n^{\ast}}}{Y_{n^{\ast}}}<\gamma^{\text{D}}_{\text{th}}| n^{\ast}=\arg\max_n  (X_n+Y_n) \right).\\
		\end{aligned}
	\end{equation} 
	Then, with the aid of the transformation of $P_{\text{DEPS}}^{\text{D}}$, an approximation of the WDT outage probability can be obtained by the following theorem.
	\begin{theorem}\label{th:SIRmaxEHP}
		The downlink WDT outage probability under the DEPS strategy can be approximated as 
		\begin{equation}\label{result:SIRmaxEHP}
			P_{\text{DEPS}}^{\text{D}}\left(\gamma^{\text{D}}_{\text{th}}\right)\approx1-\frac{1}{\left(\gamma^{\text{D}}_{\text{th}}+1\right)^{M-1}}.
		\end{equation}
	\end{theorem}
	\begin{IEEEproof}
		Please refer to Appendix \ref{proofth:SIRmaxEHP}.
	\end{IEEEproof}
	
	Next, we will work out the uplink WDT outage probability under the DEPS strategy, which is expressed as
	\begin{equation}
		\begin{aligned}
			P_{\text{DEPS}}^{\text{U}}\left(\gamma^{\text{U}}_{\text{th}}\right)&\hspace{0.5mm}=\text{Pr}\left(\gamma_{\text{U},n^{\ast}}^{(m)}<\gamma^{\text{U}}_{\text{th}}\left\vert n^{\ast}=\arg\max_{n} E_n^{(m)}\right)\right.\\
			&\hspace{0.5mm}=\text{Pr}\left(\alpha_{n^{\ast}}\beta_{n^{\ast}}<\widetilde{\gamma}\left\vert n^{\ast}=\arg\max_{n} \alpha_n \right.\right) \\
			&\overset{\left(a\right)}{=}\text{Pr}\left(\max_{n}\alpha_n <\frac{\widetilde{\gamma}}{\beta_{n^{\ast}}}\right),\\
		\end{aligned}
	\end{equation}
	where $(a)$ accounts for the fact that $\alpha_n$ is independent of $\beta_{n^{\ast}}$.
	
	Then, the uplink WDT outage probability is analyzed in the following theorem.
	\begin{theorem}\label{th:SNREPSmultipleusers}
		The uplink WDT outage probability under the DEPS
		strategy is formulated as in  \eqref{eq:SNREPSmultipleusers}.
		\begin{figure*}[]
			\begin{equation}\label{eq:SNREPSmultipleusers}
			P_{\text{DEPS}}^{\text{U}}\left(\widetilde{\gamma}\right)=\int_{x=0}^{\infty}\left(1-\mu^2\right)\exp\left[-\left(1-\mu^2\right)x\right]	\prod_{b=1}^{B}\int_{r_b=0}^{\infty}\frac{\left(r_b\right)^{M-1}\exp\left(-r_b\right)}{\Gamma(M)} \left[1-Q_M\left(\sqrt{\frac{2\mu^2r_b}{1-\mu^2}},\sqrt{\frac{\widetilde{\gamma}}{2x}}\right)\right]^{L_b}dr_bdx\\
			\end{equation}
			\hrulefill
		\end{figure*}
	\end{theorem}
	\begin{IEEEproof}
		Please refer to Appendix \ref{proofth:SNRmaxSNR}.
	\end{IEEEproof}
	
	Note that \eqref{eq:SNREPSmultipleusers} in Theorem \ref{th:SNREPSmultipleusers} contains the product term $\left[1 - Q_p(a, b)\right]^{L_b}$, which is analytically intractable due to the nested structure involving the Marcum-$Q$ function and the $L_b$-th power.  Therefore, we employ the SFA presented in Lemma~\ref{lemma:semiAppro} to simplify this expression.

	\begin{corollary}\label{lemma:EPSmultipleusers}
			As $\mu^2 \rightarrow 1$, the uplink WDT outage probability under the DEPS strategy can be further approximated by
			\begin{equation}\label{eqapproximate:EPSmultipleusers}
				P_{\text{DEPS}}^{\text{U}}\left(\gamma^{\text{U}}_{\text{th}}\right)\approx\int_{0}^{\infty}e^{-x}\prod_{b=1}^{B}\left[\frac{\Phi\left(M,  \delta_{\text{DEPS}}\left(\frac{x}{1-\mu^2},L_b\right)\right)}{\Gamma\left(M\right)}\right]dx.
			\end{equation}
			where $\Phi(a,x)=\int_{0}^{x}t^{a-1}e^{-t}dt$ is the lower incomplete gamma function, and the threshold $\delta_{\text{DEPS}}(x,L_b)$ is given by
			\begin{equation}\label{eq:DEPSthreshold}
				\begin{aligned}
					&\delta_{\text{DEPS}}\left(x,L_b\right) \\
					&= \frac{1-\mu^2}{2\mu^2} \left[ \sqrt{\frac{\widetilde{\gamma}}{2x}} + \frac{\frac{L_b - 1}{\sqrt{2\pi}} \sqrt{\frac{\widetilde{\gamma}}{2x}} + M - \frac{1}{2}}{ \frac{(L_b - 1)(M - \frac{1}{2})}{\sqrt{2\pi}} + \frac{M - \frac{1}{2}}{\sqrt{\frac{\widetilde{\gamma}}{2x}}} - \sqrt{\frac{\widetilde{\gamma}}{2x}} } \right]^2.
				\end{aligned}
			\end{equation}
		\end{corollary}
		\begin{IEEEproof}
			By substituting $a = \sqrt{\frac{2\mu^{2}r_{b}}{1-\mu^{2}}}$ and $b = \sqrt{\frac{\widetilde{\gamma}}{2x}}$ into Lemma~\ref{lemma:semiAppro}, we derive the closed-form expression of the threshold $\delta_{\text{DEPS}}(x, L_b)$, as given in \eqref{eq:DEPSthreshold}. Then, by substituting this result into \eqref{eq:SNREPSmultipleusers}, we obtain the following expression.
			\begin{multline}\label{eq:DEPSFA}
				P_{\text{DEPS}}^{\text{D}}\left(\widetilde{\gamma}\right) = \int_{x=0}^{\infty} \left(1 - \mu^2\right) e^{-\left(1 - \mu^2\right)x} \\
				\times \prod_{b=1}^{B} \int_{0}^{\delta_{\text{DEPS}}\left(x,L_b\right)} \frac{r_b^{M-1} e^{-r_b}}{\Gamma(M)} dr_b.
			\end{multline}
			The inner integral in \eqref{eq:DEPSFA} corresponds to the lower incomplete gamma function \cite[Eq. (3.351.1)]{gradshteyn2014table}. Finally, substituting this result into \eqref{eq:DEPSFA} yields the desired approximation in \eqref{eqapproximate:EPSmultipleusers}, which completes the proof.
	\end{IEEEproof}
	\begin{proposition}\label{EPSlowerbound}
		The uplink WDT outage probability under the DEPS strategy is lower bounded by
		\begin{equation}\label{DEPS-lowerbound}
			P_{\text{DEPS}}^{\text{U}}\left(\widetilde{\gamma}\right) > \int_{0}^{\infty} e^{-x} \left[ \frac{\Phi\left(M,\frac{\widehat{\gamma}}{x}\right)}{\Gamma(M)} \right]^B dx,
		\end{equation}
		where $\widehat{\gamma} = \frac{\gamma^{\text{U}}_{\text{th}} t_2\sigma_w^2\Omega^{2}}{4\eta(1-\rho^{(m)})P_t t_1} $.
	\end{proposition}
	
	\begin{IEEEproof}
		Please refer to Appendix~\ref{proof:EPSlowerbound}.
	\end{IEEEproof}
	\begin{proposition}\label{DEPSlowerbound-proposition2}
		The uplink WDT outage probability under the DEPS strategy is further lower bound in closed-form as
		\begin{equation}\label{DEPSlowerbound-closed}
			P_{\text{DEPS}}^{\text{U}}\left(\widehat{\gamma}\right) > \sum_{b=0}^{BM} (-1)^b \sqrt{b \widehat{\gamma} d_M} K_1\left( \sqrt{b \widehat{\gamma} d_M} \right),
		\end{equation}
		where \( d_M = \left[\Gamma\left(1+M\right)\right]^{-1/M} \).
	\end{proposition}
	
	\begin{IEEEproof}
		According to \cite{AlzerHorst}, the following lower bound for the normalized lower incomplete gamma function holds for all \( x \geq 0 \) and \( M > 1 \)
		\begin{equation}
			\frac{\Phi\left(M,x\right)}{\Gamma\left(M\right)} > \left(1 - e^{-d_M x} \right)^M,
		\end{equation}
		where \( d_M = \left[ \Gamma(1 + M) \right]^{-1/M} \). Applying this bound to the integrand in \eqref{DEPS-lowerbound}, we obtain
		\begin{equation}
			\left[ \frac{\Phi\left(M,\frac{\widehat{\gamma}}{x}\right)}{\Gamma(M)} \right]^B > \left[ 1 - e^{-d_M \frac{\widehat{\gamma}}{x}} \right]^{BM}.
		\end{equation}
		Substituting into \eqref{DEPS-lowerbound}, we arrive at
		\begin{equation}
			P_{\text{DEPS}}^{\text{U}}\left(\widetilde{\gamma}\right) > \int_{0}^{\infty} e^{-x} \left[ 1 - e^{-d_M \frac{\widehat{\gamma}}{x}} \right]^{BM} dx.
		\end{equation}
		Next, by applying the binomial expansion and the integral identity in \cite[Eq. (3.471.9)]{gradshteyn2014table}, the integral can be expressed in closed form as the summation in \eqref{DEPSlowerbound-closed}, which completes the proof.
	\end{IEEEproof}
	
	\subsection{Outage Probabilities under the
		UCPS Strategy}
	In this section, we will analyze the outage performance under the UCPS strategy. By applying the UCPS  strategy, the receiver switches the port based solely on the  uplink channel $h_n^{(m)}$, selecting the port that has the best uplink channel gain. When the user $m$ applies the UCPS strategy, it selects the optimal antenna port $n^{\ast}$ having the maximum $\vert h_n^{(m)}\vert^2$ as
	\begin{equation}\label{UPS-multi-user}
		n^{\ast}=\arg\max_{n} \vert h_n^{(m)}\vert^2.
	\end{equation}
	Thus, the downlink WDT outage probability is expressed as
	\begin{equation}\label{eqfirst:SIRmaxh}
		\begin{aligned}
			P_{\text{UCPS}}^{\text{D}}(\gamma_{\text{th}}^{\text{D}})&=\text{Pr}\left(\gamma_{\text{D},n^{\ast}}^{(m)}<\gamma_{\text{th}}^{\text{D}} \vert n^{\ast}=\arg\max_{n}\vert h_{n}^{(m)}\vert^2\right)\\
			&= \text{Pr}\left(\frac{X_{n^{\ast}}}{Y_{n^{\ast}}}<\gamma_{\text{th}}^{\text{D}}\vert n^{\ast}=\arg\max_{n}  \beta_{n} \right).\\
		\end{aligned}		
	\end{equation}
	\begin{theorem}\label{th:SIRUPSmultipleusers}
		The downlink WDT outage probability under the UCPS
		strategy is derived as
		\begin{equation}\label{eq:SIRUPSmultipleusers}
			P_{\text{UCPS}}^{\text{D}}(\gamma_{\text{th}}^{\text{D}})=1-\frac{1}{\left(\gamma_{\text{th}}^{\text{D}}+1\right)^{M-1}}.
		\end{equation}
	\end{theorem}
	\begin{IEEEproof}
		Due to the independence between the uplink and downlink channels, selecting the port based on the uplink channel is equivalent to the outage probability in single antenna  scenario. Thus, \eqref{eqfirst:SIRmaxh} is then simplified as
		\begin{equation}\label{eqsimplify:SIRmaxh}
			P_{\text{UCPS}}^{\text{D}}(\gamma_{\text{th}}^{\text{D}})
			=\text{Pr}\left(\frac{X_{n^{\ast}}}{Y_{n^{\ast}}}<\gamma_{\text{th}}^{\text{D}}\right), \forall n^{\ast}.
		\end{equation}
		Similar to the case of DEPS strategy,  \eqref{eqsimplify:SIRmaxh} is further derived, and the proof ends.
	\end{IEEEproof}
	
	In what follows, we will analyze the uplink WDT outage probability, which is expressed as
	\begin{equation}\label{eqfirstSNRmaxh}
		\begin{aligned}
			P_{\text{UCPS}}^{\text{U}}\left(\gamma_{\text{th}}^{\text{U}}\right)&=\text{Pr}\left(\gamma_{\text{U},n^{\ast}}^{(m)}<\gamma_{\text{th}}^{\text{U}}\left\vert n^{\ast}=\arg\max_{n} \vert h_n^{(m)}\vert^2\right)\right.\\
			&=\text{Pr}\left(\alpha_{n^{\ast}}\beta_{n^{\ast}}<\widetilde{\gamma}\left\vert n^{\ast}=\arg\max_{n} \beta_n \right.\right) \\
			&=\text{Pr}\left(\max_{n}\beta_n <\frac{\widetilde{\gamma}}{\alpha_{n^{\ast}}}\right).
		\end{aligned}
	\end{equation}
	\begin{theorem}\label{th:SNRUPSmultipleusers}
		The uplink WDT outage probability under the UCPS
		strategy is expressed as in \eqref{eq:SNRUPSmultipleusers}.
		\begin{figure*}[]
			\begin{equation}\label{eq:SNRUPSmultipleusers}
				P_{\text{UCPS}}^{\text{U}}\left(\widetilde{\gamma}\right)=\frac{\left(1-\mu^2\right)^{M}}{\Gamma(M)}\int_{y=0}^{\infty}\exp\left(-y+\mu^2y\right)y^{M-1}dy\prod_{b=1}^B{\int_{\widetilde{r}_b=0}^{\infty}e^{-\widetilde{r}_b} \left[1-Q_{1}\left(\sqrt{\frac{2\mu^{2}\widetilde{r}_b}{1-\mu^{2}}},\sqrt{\frac{\widetilde{\gamma}}{2y}}\right)\right]^{L_b}d\widetilde{r}_b}\\
			\end{equation}
			\hrulefill
		\end{figure*}
	\end{theorem}
	\begin{IEEEproof}
		Please refer to Appendix \ref{prooth:SNRUPSmultipleusers}.
	\end{IEEEproof}6
	
	Similar to the case in Theorem~\ref{th:SNREPSmultipleusers}, the expression in \eqref{eq:SNRUPSmultipleusers} contains $B$ inner integrals nested within an outer integral, making it analytically intractable and computationally intensive.  To address this challenge, we adopt the SFA proposed in Lemma~\ref{lemma:semiAppro}, which enables a more tractable and computationally efficient formulation for further analysis.
	
		\begin{corollary}\label{lemma:UPSmultipleusers}
			As $\mu^2 \rightarrow 1$, the uplink WDT outage probability under the UCPS strategy can be approximated as
			\begin{multline}\label{eq:approximate:UPSmultipleusers}
				P_{\text{UCPS}}^{\text{U}}\left(\widetilde{\gamma}\right)\approx\frac{1}{\Gamma(M)}\int_{y=0}^{\infty}e^{-y}y^{M-1}\\\times\prod_{b=1}^{B}\left[1-e^{-\delta_{\text{UCPS}}\left(\frac{y}{1-\mu^2},L_b\right)}\right]dy.
			\end{multline}
			where the threshold is 
				\begin{equation}\label{eq:UCPSthreshold}
				\delta_{\text{UCPS}}\left(y,L_b\right) = \frac{1-\mu^2}{2\mu^2} \left[ \sqrt{\frac{\widetilde{\gamma}}{2y}} + \frac{\frac{L_b - 1}{\sqrt{2\pi}} \sqrt{\frac{\widetilde{\gamma}}{2y}} +  \frac{1}{2}}{ \frac{(L_b - 1) }{2\sqrt{2\pi}} + \frac{ \frac{1}{2}}{\sqrt{\frac{\widetilde{\gamma}}{2y}}} - \sqrt{\frac{\widetilde{\gamma}}{2y}} } \right]^2.
			\end{equation}
		\end{corollary}
		
		\begin{IEEEproof}
			By substituting \(a = \sqrt{\frac{2\mu^2 r_b}{1 - \mu^2}}\) and \(b = \sqrt{\frac{\widetilde{\gamma}}{2y}}\) into Lemma~\ref{lemma:semiAppro}, the threshold \(\delta_{\text{UCPS}}(y,L_b)\) can be obtained as in \eqref{eq:UCPSthreshold}.
			Based on this approximation, the uplink outage probability in \eqref{eq:SNRUPSmultipleusers} can be rewritten as
			\begin{equation}
				\begin{aligned}
					P_{\text{UCPS}}^{\text{U}}\left(\widetilde{\gamma}\right)&\approx \frac{(1 - \mu^2)^M}{\Gamma(M)}\times\\ 
					& \int_{0}^{\infty} e^{-(1 - \mu^2) y} y^{M-1} \prod_{b=1}^{B} \int_{r_b=0}^{\delta_{\text{UCPS}}\left(y,L_b\right)} e^{-r_b} dr_bdy.\\
				\end{aligned}
			\end{equation}
			Therefore, the approximate expression in \eqref{eq:approximate:UPSmultipleusers} can be directly obtained, which completes the proof.
	\end{IEEEproof}
	The approximation in \eqref{eq:approximate:UPSmultipleusers} leverages the SFA to transform the original nested integral structure into a single integral, significantly reducing the computational complexity.
	\begin{proposition}\label{Pro:SNRmaxG-closedform}
		The uplink WDT outage probability  under UCPS strategy is lower bounded by
		\begin{equation}\label{lowerboundSNRmaxH}
			P_{\text{UCPS}}^{\text{U}}\left(\widetilde{\gamma}\right)>\frac{1}{2^{M-1}\Gamma\left(M\right)}\sum_{b=0}^B\left(-1\right)^b\left(\sqrt{b\widetilde{\gamma}}\right)^MK_M\left(\sqrt{b\widetilde{\gamma}}\right).
		\end{equation}
	\end{proposition}
	\begin{IEEEproof}
		The proof follows a similar procedure to that of Proposition~\ref{EPSlowerbound} and is thus omitted for brevity.
	\end{IEEEproof}
	\subsection{Outage Probabilities under the USPS Strategy}
	In this section, we study a port selection strategy based on the received SNR at the HAP from  the user $m$ , referred to as the USPS strategy. This strategy requires knowledge of both the downlink signals received by user $m$ and the transmitted uplink signal. If the USPS strategy is applied, the user selects the optimal antenna port $n^{\ast}$ that achieves the maximum of the received SNR at the HAP, which is expressed as
	\begin{equation}
		n^{\ast}=\arg\max_{n} 	\gamma^{(m)}_{\text{U},n}=\arg\max_{n}\vert h_n^{(m)}\vert^2\sum_{\tilde{m}=1}^M\vert g_n^{(\tilde{m},m)}\vert^2.
	\end{equation}
	Firstly, the downlink WDT outage probability based on the USPS strategy is expressed as
	\begin{equation}
		\begin{aligned}
			P_{\text{USPS}}^{\text{D}}(\gamma_{\text{th}}^{\text{D}})&=\text{Pr}\left(\gamma^{(m)}_{\text{D},n^{\ast}}<\gamma_{\text{th}}^{\text{D}} \left\vert n^{\ast}=\arg\max_{n}   \gamma^{(m)}_{\text{U},n} \right.\right)\\
			&=\text{Pr}\left(\frac{X_{n^{\ast}}}{Y_{n^{\ast}}}<\gamma_{\text{th}}^{\text{D}} \left\vert n^{\ast}=\arg\max_{n}    \beta_n\alpha_n\right.\right),
		\end{aligned}		
	\end{equation}
	Next, we investigate the downlink WDT outage probability in the following theorem.
	\begin{theorem}\label{th:SIRmaxSNR}
		The downlink WDT outage probability  by applying the USPS strategy is approximated as 
		\begin{equation}\label{eq:SIRmaxSNR}
			P_{\text{USPS}}^{\text{D}}\left(\gamma_{\text{th}}^{\text{D}}\right)\approx1-\frac{1}{\left(\gamma_{\text{th}}^{\text{D}}+1\right)^{M-1}}.
		\end{equation}
	\end{theorem}
	\begin{IEEEproof}
		Since $\mu^2$ is close to 1, using the result of Lemma \ref{lemma:XYindependent}, $X_{n^{\ast}}/Y_{n^{\ast}}$, $\alpha_n$ and $\beta_n$ becomes independent to each other. Then, the downlink WDT outage probability under the USPS strategy can be approximated by that of a single traditional antenna as
		\begin{equation}
			P_{\text{USPS}}^{\text{D}}\left(\gamma_{\text{th}}^{\text{D}}\right)\approx \text{Pr}\left(\gamma_{\text{D},n^{\ast}}^{(m)}=\frac{X_{n^{\ast}}}{Y_{n^{\ast}}}<\gamma_{\text{th}}^{\text{D}}\right),\forall n^{\ast}.
		\end{equation}
		After that, \eqref{eq:SIRmaxSNR} can be obtained by following the same steps as the proof of Theorem \ref{th:SIRmaxEHP}.
	\end{IEEEproof}
	
	Then, we aim to study the uplink WDT outage probability under the USPS strategy, which is expressed as
	\begin{equation}
		\begin{aligned}
			P_{\text{USPS}}^{\text{U}}(\gamma_{\text{th}}^{\text{U}})
			&=\text{Pr}\left(\arg\max_{n} \gamma_{\text{U},n}^{(m)}<\gamma_{\text{th}}^{\text{U}}\right)\\
			&=\text{Pr}\left(\arg\max_{n} \alpha_n\beta_n<\widetilde{\gamma}\right).
		\end{aligned}		
	\end{equation}
	After that, we investigate the uplink WDT outage probability in the following theorem.
	\begin{theorem}\label{th:SNRmaxSNR}
		The uplink WDT outage probability under the USPS strategy is formulated as in \eqref{eq:SPS-multipleusers}. 
		\begin{figure*}[]
			\begin{equation}\label{eq:SPS-multipleusers}
				P_{\text{USPS}}^{\text{U}}\left(\widetilde{\gamma}\right)=\prod_{b=1}^{B}\int_{0}^{\infty}\int_{0}^{\infty}\frac{r_{b}^{M-1}e^{-\widetilde{r}_{b}-r_b}}{\Gamma(M)}\left[1-\int_{z=0}^{\infty}Q_{M}\left(\sqrt{\frac{2\mu^{2}r_{b}}{1-\mu^{2}}},\sqrt{\frac{\widetilde{\gamma}}{2z}}\right)e^{-z-\frac{\mu^2\widetilde{r}_b}{1-\mu^2}}I_0\left(2\sqrt{\frac{\mu^2\widetilde{r}_bz}{1-\mu^2}}\right)dz\right]^{L_b}dr_bd\widetilde{r}_{b}\\
			\end{equation}
			\hrulefill
		\end{figure*}
	\end{theorem}
	\begin{IEEEproof}
		The uplink WDT outage probability under the USPS strategy is formulated as in \eqref{SPSproofprocess-multi-user}, where $\left(a\right)$ uses the results of \eqref{jointCDFalphan} and  \eqref{PDFbetanast} as well as the variable substitutions. Then, \eqref{eq:SPS-multipleusers} is obtained, which completes the proof.
		\begin{figure*}[t]
			\begin{equation}\label{SPSproofprocess-multi-user}
				\begin{aligned}
					&P_{\text{USPS}}^{\text{U}}\left(\gamma^{\text{U}}_{\text{th}}\right)=P\left(\arg\max_{n} \alpha_n\beta_n<\widetilde{\gamma}\right)=\int_{0}^{\infty}\ldots\int_{0}^{\infty}F_{\alpha_n\vert \beta_n}\left(\frac{\widetilde{\gamma}}{y_1},\ldots,\frac{\widetilde{\gamma}}{y_N}\right)f_{\beta_n}\left(y_1,\ldots,y_N\right)dy_1,\ldots,dy_N\\
					&\overset{\left(a\right)}{=}\prod_{b=1}^{B}\int_{0}^{\infty}\int_{0}^{\infty}\frac{r_{b}^{M-1}e^{-\frac{\widetilde{r}_{b}+r_b}{2}}}{2^{M+1}\Gamma(M)}\left[\int_{y=0}^{\infty}\left(1-Q_{M}\left(\sqrt{\frac{\mu^{2}r_{b}}{1-\mu^{2}}},\sqrt{\frac{\widetilde{\gamma}}{y}}\right)\right)\frac{1}{2}\exp\left[-\frac{y+\frac{\mu^2\widetilde{r}_b}{1-\mu^2}}{2}\right]I_0\left(\sqrt{\frac{\mu^2\widetilde{r}_by}{1-\mu^2}}\right)dy\right]^{L_b}dr_bd\widetilde{r}_{b}\\
				\end{aligned}		
			\end{equation}
			\hrulefill
		\end{figure*}
	\end{IEEEproof}
	
	\begin{corollary}\label{lemma:SPSSFAmultipleusers}
			The uplink WDT outage probability under the USPS strategy can be further approximated as
			\begin{equation}\label{SPSSFA-multipleusers}
				\begin{aligned}
					P_{\text{USPS}}^{\text{U}}\left(\widetilde{\gamma}\right)\approx\prod_{b=1}^B\left[1-\int_{r_b=0}^{\infty}\frac{r_{b}^{M-1}e^{-r_b-\tilde{\delta}_{\text{USPS}}\left(r_b,L_b\right)}}{\Gamma(M)}dr_b\right],
				\end{aligned}		
			\end{equation}
			where
			\begin{equation}\label{delta0USPS}
				\begin{aligned}
					&\tilde{\delta}_{\text{USPS}}(r_b, L_b)  \\
					&=\frac{1 - \mu^2}{2\mu^2}
					\left[
					\delta_{\text{USPS}}(r_b) +
					\frac{
						\frac{L_b - 1}{\sqrt{2\pi}} \delta_{\text{USPS}}(r_b) + \frac{1}{2}
					}{
						\frac{L_b - 1}{2\sqrt{2\pi}} + \frac{1}{2\delta_{\text{USPS}}(r_b)} - \delta_{\text{USPS}}(r_b)
					}
					\right]^2.
				\end{aligned}
			\end{equation}
			and
			\begin{equation}\label{delta1USPS}
				\delta_{\text{USPS}}\left(r_b\right)=\frac{2\sqrt{\widetilde{\gamma}}}{\sqrt{\frac{2\mu^2r_b}{1-\mu^2}}+\sqrt{\frac{2\mu^2r_b}{1-\mu^2}+4M-2}}.
			\end{equation}
		\end{corollary}
		\begin{IEEEproof}
			With the aid of SFA in Lemma \ref{lemma:semiApproL=1}, \eqref{eq:SPS-multipleusers} is re-formulated  as
			\begin{multline}\label{SNRmaxSNRprocess1}
				P_{\text{USPS}}^{\text{U}}\left(\widetilde{\gamma}\right)
				\approx
				\prod_{b=1}^B\int_{0}^{\infty}\int_{0}^{\infty}\frac{r_{b}^{M-1}e^{-\widetilde{r}_{b}-r_b}}{\Gamma(M)}\left[1-\int_{\delta_{\text{USPS}}\left(r_b\right)}^{\infty}\right.\\
				\left.e^{-y-\frac{\mu^2\widetilde{r}_b}{1-\mu^2}} I_0\left(2\sqrt{\frac{\mu^2\widetilde{r}_by}{1-\mu^2}}\right)dy\right]^{L_b}dr_bd\widetilde{r}_b. 
			\end{multline}
			Note that the generalized Marcum-$Q$-function can be alternatively defined as a finite integral. Thus, \eqref{SNRmaxSNRprocess1} can be rewritten as
			\begin{equation}\label{SNRmaxSNRprocess2}
				\begin{aligned}
					P_{\text{USPS}}^{\text{U}}\left(\widetilde{\gamma}\right)=&	\prod_{b=1}^B\int_{0}^{\infty}\int_{0}^{\infty}\frac{r_{b}^{M-1}e^{-\widetilde{r}_{b}-r_b}}{\Gamma(M)}	\times\\
					&\left[1-	Q_1\left(\sqrt{\frac{2\mu^2\widetilde{r}_b}{1-\mu^2}},\delta_{\text{USPS}}\left(r_b\right)\right)\right]^{L_b}dr_bd\widetilde{r}_b.
				\end{aligned}		
			\end{equation}
			where $	\delta_{\text{USPS}}\left(r_b\right)$ is given in \eqref{delta1USPS}.
			Then, due to the presence of the $\left[1-Q_p\left(a,b\right)\right]^{L_b}$, we can apply the SFA once again. Thus, the new threshold  $\tilde{\delta}_{\text{USPS}}(r_b,L_b)$ is obtained as \eqref{delta0USPS}. After that, \eqref{SNRmaxSNRprocess2} is approximated by
			\begin{equation}\label{SPSSFAprocess-multipleusers}
				\begin{aligned}
					P_{\text{USPS}}^{\text{U}}\left(\widetilde{\gamma}\right)\approx\prod_{b=1}^B\int_{r_b=0}^{\infty}\frac{r_{b}^{M-1}e^{-r_b}}{\Gamma(M)}\int_{\widetilde{r}_b=0}^{\tilde{\delta}_{\text{USPS}}\left(r_b,L_b\right)}e^{-\widetilde{r}_b}d\widetilde{r}_bdr_b.
				\end{aligned}		
			\end{equation}
			Finally, \eqref{SPSSFA-multipleusers} is obtained, which completes the proof.
	\end{IEEEproof}
	\begin{proposition}\label{SPSlowerbound}
		The uplink WDT outage probability under USPS strategy is lower bounded as
		\begin{equation}\label{USPS-lowerbound}
			P_{\text{USPS}}^{\text{U}}(\widehat{\gamma})>\left[1-\frac{\left(\sqrt{\widehat{\gamma}}\right)^M}{2^{M-1}\Gamma\left(M\right)}K_M\left(\sqrt{\widehat{\gamma}}\right)\right]^{B}.\\
		\end{equation}
	\end{proposition}
	\begin{IEEEproof}
		Please refer to Appendix \ref{proof:SPSlowerbound}.
	\end{IEEEproof}
	\subsection{Further discussion}
	
		According to~\cite[Eq. (20)]{10103838}, under the condition
		$0 < \epsilon < \frac{N-1}{W},$
		the number of eigenvalues $B$ exceeding the threshold \(\epsilon\) can be approximately estimated as $	B \approx 2W  \frac{N}{N-1}$, for large $N$.  Therefore, with a fixed antenna size $W$ and increasing port number $N$, the outage probability initially decreases and then converges due to the limitation imposed by $W$. Similarly, with a fixed $N$ and increasing $W$, the uplink WDT outage probability decreases as the distance between ports increases, eventually converging to the derived lower bound when the wireless channels among ports become independent.
		
		Moreover, Theorem~\ref{th:SNRWPSmultipleusers} provides an approximate closed-form expression for the uplink WDT outage probability under the DSPS strategy. Theorems~\ref{th:SIRmaxEHP}, \ref{th:SIRUPSmultipleusers}, and \ref{th:SIRmaxSNR} present closed-form expressions for the downlink WDT outage probabilities under the DSPS, UCPS, and USPS strategies, respectively. Notably, all these results are independent of $N$ and $W$, indicating that increasing the antenna size or the number of ports does not yield performance gains in these scenarios.

	\section{Numerical Results}\label{umerical Results}
	This section presents results obtained through Gauss-Laguerre quadrature (GLQ) \cite{abramowitz1968handbook} for the theoretical expressions in Theorems \ref{th:SIRWPSmultipleusers}, \ref{th:SNREPSmultipleusers}, \ref{th:SNRUPSmultipleusers}, and \ref{th:SNRmaxSNR}, along with results derived using the SFA in Corollaries \ref{corollary:SIRmaxSIRSFAtwice}, \ref{lemma:EPSmultipleusers}, \ref{lemma:UPSmultipleusers}, and \ref{lemma:SPSSFAmultipleusers}. We also discuss the approximations in Theorems \ref{th:SNRWPSmultipleusers}, \ref{th:SIRmaxEHP}, and \ref{th:SIRmaxSNR}, and include lower bounds from Propositions \ref{EPSlowerbound}, \ref{Pro:SNRmaxG-closedform}, and \ref{SPSlowerbound} for a comprehensive analysis. Monte Carlo simulations are used to validate the theoretical results, evaluating the outage performance of FAMA-assisted WPCN. A summary of simulation-based performance metrics is provided in Table \ref{tab:strategies}. The simulation parameters are as follows: The energy conversion efficiency is set to $\eta = 0.45$, and the large-scale path loss is modeled as $\Omega = 10^{-3} d^{-\zeta}$, where $d$ denotes the distance between the HAP antenna and the user, and $\zeta = 2.2$ is the path loss exponent \cite{10422762}. Note that the constant $10^{-3}$ corresponds to a reference signal attenuation of 30 dB at a distance of 1 meter. In the Monte Carlo simulations, multiple independent realizations of Rayleigh fading channels are generated according to \eqref{DLchannel} and \eqref{ULchannel}. The outage performance is then evaluated by averaging over these independent channel realizations.The variance of the additive white Gaussian noise (AWGN) is set to $\sigma_w^2 = -90$ dBm~\cite{9928052}. The time allocation between the IDET and WDT phases is 4:1, and the correlation coefficient is set to $\mu^2 = 0.97$ unless otherwise specified. Other parameter settings are detailed in each figure.
	\begin{table*}[t]
		\centering
		\caption{Outage Performance Metrics Summary}
		\label{tab:strategies}
		\resizebox{0.9\linewidth}{!}{%
			\begin{tabular}{@{}lcccc@{}}
				\toprule
				& \textbf{DSPS stategy} & \textbf{DEPS stategy} & \textbf{UCPS stategy} & \textbf{USPS stategy} \\ \midrule
				\multirow{2}{*}{\textbf{Downlink WDT}} & Theorem \ref{th:SIRWPSmultipleusers} (GLQ) & \multirow{2}{*}{Theorem \ref{th:SIRmaxEHP} (Appro.)} & \multirow{2}{*}{Theorem \ref{th:SIRUPSmultipleusers} (Ana.)}& \multirow{2}{*}{Theorem \ref{th:SIRmaxSNR} (Appro.)} \\ 
				&Corollary \ref{corollary:SIRmaxSIRSFAtwice} (SFA) & &   &   \\
				\midrule
				\multirow{3}{*}{\textbf{Uplink WDT}}   & \multirow{3}{*}{Theorem \ref{th:SNRWPSmultipleusers} (Appro.)}&Theorem \ref{th:SNREPSmultipleusers} (GLQ) & Theorem \ref{th:SNRUPSmultipleusers} (GLQ) &Theorem  \ref{th:SNRmaxSNR} (GLQ) \\ 
				&  &Corollary \ref{lemma:EPSmultipleusers} (SFA) & Corollary \ref{lemma:UPSmultipleusers} (SFA) & Corollary \ref{lemma:SPSSFAmultipleusers} (SFA)\\
				&  &Proposition \ref{EPSlowerbound} (Lower bound)& Proposition \ref{Pro:SNRmaxG-closedform} (Lower bound) & Proposition \ref{SPSlowerbound} (Lower bound)\\ \bottomrule
		\end{tabular}}
	\end{table*}
	
	\begin{figure*}[htbp]
		\centering
		\begin{minipage}[t]{0.3\textwidth}
			\centering
			\includegraphics[width=\textwidth]{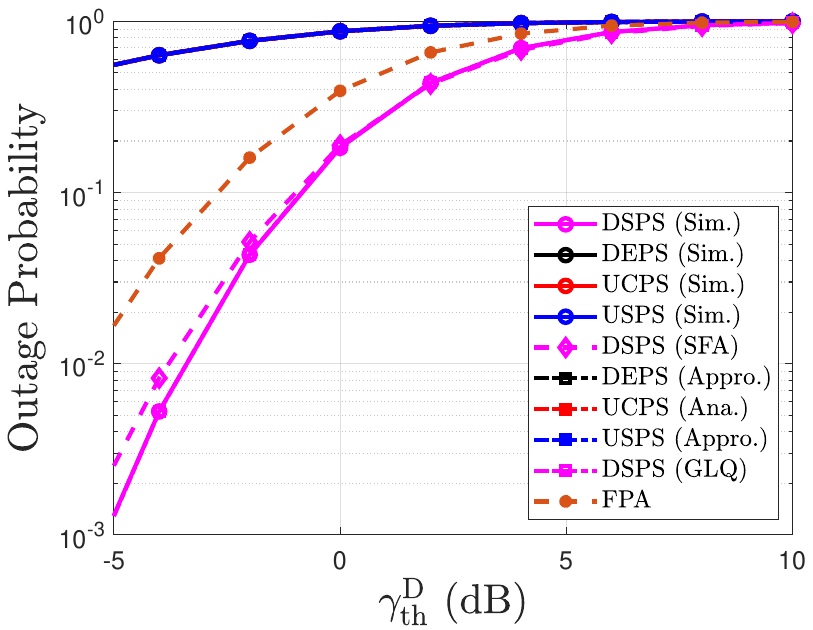}  
			\caption{Downlink WDT outage probability versus $\gamma_{\text{th}}^{\text{D}}$;  $M = 4$, $N=50$, $W=3$.}  
			\label{SIRvsGamma}
		\end{minipage} \hfill
		\begin{minipage}[t]{0.3\textwidth}
			\centering
			\includegraphics[width=\textwidth]{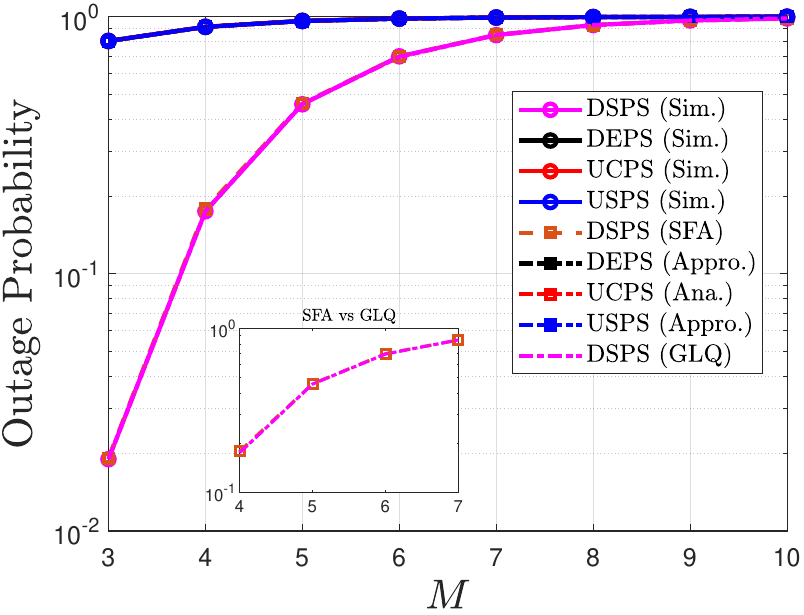}  
			\caption{ Downlink WDT outage probability versus $M$; $\gamma_{\text{th}}^{\text{D}} = 1$ dB, $N=50$, $W=5$.}  
			\label{SIRvsU}
		\end{minipage} \hfill
		\begin{minipage}[t]{0.3\textwidth}
			\centering
			\includegraphics[width=\textwidth]{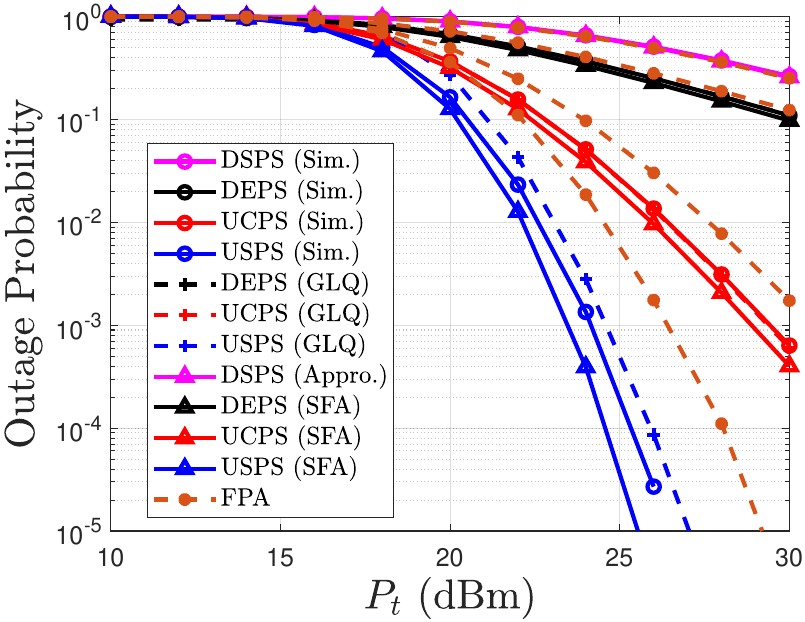}  
			\caption{Uplink WDT outage probability versus transmit power $P_t$; $N=50$, $M=4$, $W=4$, $\gamma_{\text{U,th}}^{(m)}=10$ dB, $d=18$.}  
			\label{Pt-multiuser}
		\end{minipage} \\[1ex] 
		
		\begin{minipage}[t]{0.3\textwidth}
			\centering
			\includegraphics[width=\textwidth]{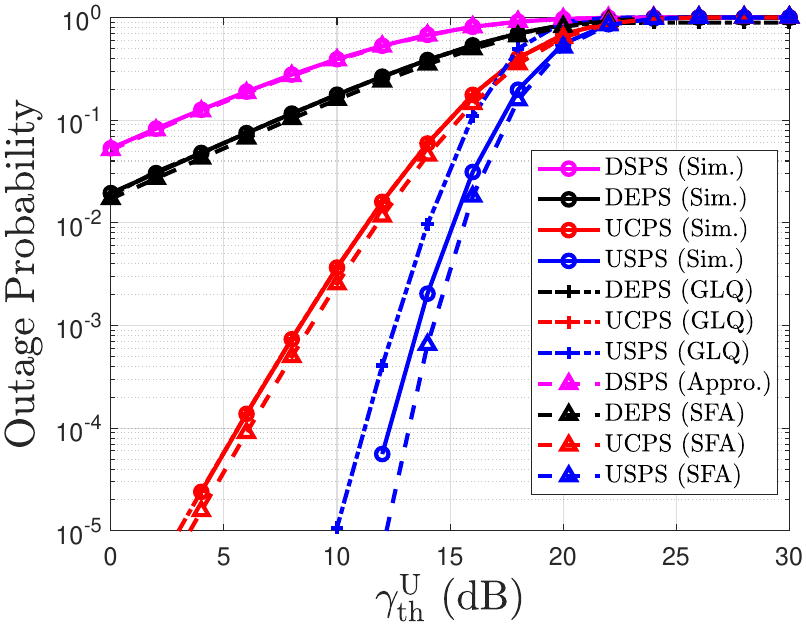}  
			\caption{Uplink WDT outage probability versus uplink SNR threshold $\gamma_{\text{th}}^{\text{U}}$;  $P_t=20$ dBm, $N=50$, $M=4$, $W=4$, $d=12$.}  
			\label{gamma-multiuser}
		\end{minipage} \hfill
		\begin{minipage}[t]{0.3\textwidth}
			\centering
			\includegraphics[width=\textwidth]{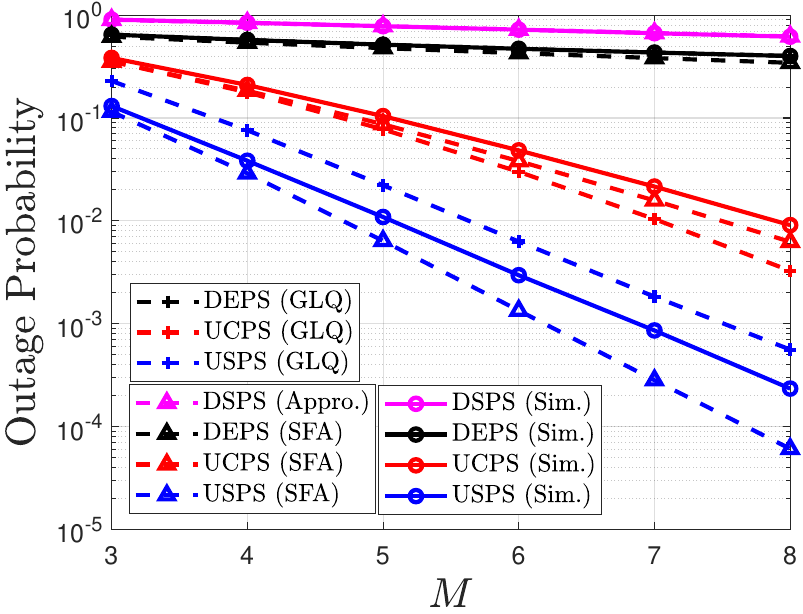}  
			\caption{Uplink WDT outage probability versus $M$; $P_t=20$ dBm, $\gamma_{\text{th}}^{\text{U}}=5$ dB, $N=50$, $W=5$, $d=22$.}  
			\label{U-multiuser}
		\end{minipage} \hfill
		\begin{minipage}[t]{0.3\textwidth}
			\centering
			\includegraphics[width=\textwidth]{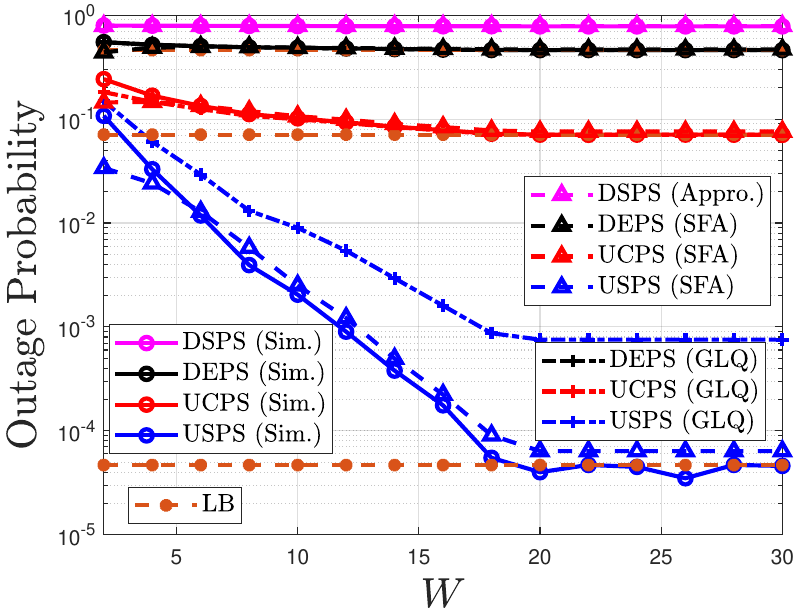}  
			\caption{Uplink WDT outage probability versus $W$; $M = 4$, $N=50$, $P_t=25$ dBm, $\gamma_{\text{th}}^{\text{U}}=10$ dB, $d=21$.}  
			\label{SNROP-W}
		\end{minipage}
	\end{figure*}
	\subsection{Downlink WDT outage probability}
	As shown in the \cref{SIRvsGamma,SIRvsU}, the outage probability increases with the growth of $M$ or $\gamma_{\text{th}}^{\text{D}}$, as expected. The downlink WDT outage performance under the DSPS strategy surpasses that of the USPS, DEPS, and UCPS strategies due to the DSPS strategy’s ability to select the port with the least interference.  Moreover, the theoretical analysis for DSPS by using GLQ exhibits excellent agreement with the simulation results. The approximate closed-form expressions for the outage probabilities under the USPS and DEPS strategies closely match the simulated values, confirming the accuracy of the approximations. The downlink WDT outage probability derived in Theorem \ref{th:SIRUPSmultipleusers} perfectly aligns with the simulated data, thereby validating the theoretical derivation. 
	
	 Additionally, under the DSPS strategy, we apply the SFA twice to derive a single-integral form expression, as presented in Corollary~\ref{corollary:SIRmaxSIRSFAtwice}. Fig.~\ref{SIRvsGamma} shows that although this approximation is less accurate for small \(\gamma_{\text{th}}^{\text{D}}\) (\textit{e.g.,} \(\gamma_{\text{th}}^{\text{D}} < 0\)), it matches the simulation results well for large \(\gamma_{\text{th}}^{\text{D}}\) (\textit{e.g.,} \(\gamma_{\text{th}}^{\text{D}} \geq 0\)). This is reasonable, as analyzed in Section~\ref{SFAmethod}. When $\gamma_{\text{th}}^{\text{D}}$ increases, the product of the Marcum-$Q$ function parameters $a$ and $b$ becomes larger, resulting in a more accurate approximation. It also  effectively avoids the need for computing multiple integrals. As shown in Fig.~\ref{SIRvsU}, the approximate outage probabilities under the given setting closely match the simulation results, further confirming the accuracy and effectiveness of the approximation method.
	
Finally, the WDT outage performance of the proposed FAMA scheme is compared with that of a conventional FPA system. As the FA system effectively degenerates into a single-antenna system under other port selection strategies, the comparison is performed only under the DSPS strategy. The benchmark is the selection combining system with the maximum
number of uncorrelated antennas allowed in the given space. For instance, when the FA size is $W = 1$, the benchmark corresponds to an SC system with 3 uncorrelated antennas; when $W = 3$, it includes 7 antennas. Simulation results demonstrate that, under identical spatial constraints and with the DSPS strategy, a single FA can outperform the conventional SC-based system, highlighting the superior spatial efficiency and performance of the FAMA architecture.

	\subsection{Uplink WDT outage probability}
	\cref{Pt-multiuser} illustrates the uplink WDT outage probability versus transmit power $P_t$ under the four port selection strategies. The results obtained using GLQ closely match the simulation outcomes, validating the correctness of the theoretical analysis. The results obtained using the SFA under the DEPS, USPS, and UCPS strategies closely match the simulation values, demonstrating the effectiveness of the approximation method. As expected, the uplink WDT outage probability decreases as $P_t$ increases for all four strategies. It can also be observed that among the four proposed port selection strategies, the DSPS strategy exhibits the poorest uplink WDT performance, while the USPS strategy achieves the best performance. Interestingly, although the DEPS strategy selects the port that harvests the most EHP, the uplink WDT outage probability is worse than the UCPS strategy, which selects the port based solely on the uplink channel. Similarly, the uplink WDT performance of the FAMA architecture is compared with that of the conventional FPA system. The results show that, when the antenna size is $W = 4$ and the number of ports is $N = 50$, the FA outperforms a traditional SC system with 9 fixed antennas under both the USPS and UCPS strategies, with particularly notable gains under the USPS strategy. Under the DEPS strategy, the FA performs slightly better, while under the DSPS strategy, both systems exhibit identical performance since the FA effectively degenerates into a single-antenna system in this case.
	
	\cref{gamma-multiuser} shows the effect of the uplink SNR threshold on the uplink WDT outage probability. The performance under these four strategies is consistent with those shown in \cref{Pt-multiuser}, \textit{i.e.,} USPS strategy performs the best compared to the other three strategies. Apparently, as $\gamma_{\text{th}}^{\text{U}}$ increases, the uplink WDT outage probability  also increases, which agrees with the Theorem \ref{th:SNRWPSmultipleusers},
	Corollaries \ref{lemma:EPSmultipleusers}, \ref{lemma:UPSmultipleusers}, and \ref{lemma:SPSSFAmultipleusers}. It is worth noting that under the UCPS strategy, the results obtained using the SFA closely match the simulation values. This is due to the suitability of the approximation, since a smaller order ($p=1$) of the Marcum-$Q$ function enhances the accuracy of the approximation, as analyzed in Section~\ref{SFAmethod}. 
	
	\cref{U-multiuser} is provided to further discuss the effect of $M$ on the uplink WDT performance. It can be observed that as the number of users $M$ increases, the uplink outage probability decreases. This is because more users result in more energy being collected, leading to a better performance, which agrees with the Theorem \ref{th:SNRWPSmultipleusers},
	Corollaries \ref{lemma:EPSmultipleusers}, \ref{lemma:UPSmultipleusers}, and \ref{lemma:SPSSFAmultipleusers}. Besides, under the DSPS and DEPS strategies, the improvement in performance with an increasing number of users is much smaller compared to the other two strategies. This is because the UCPS and USPS strategies directly impact the uplink WDT performance, whereas the DSPS and DEPS strategies influence the uplink WDT performance indirectly by affecting the amount of energy harvested during the downlink IDET phase. Moreover, as $M$ increases, the approximation error also grows under UCPS and USPS strategies. Specifically, for the USPS strategy, the error is larger than that of UCPS because it involves applying the approximation twice, which leads to error accumulation.

	\cref{SNROP-W} explores the  impact of antenna size  $W$\footnote{Note that large antenna size \textit{i.e.,} $W=30$, is not be feasible for user-side deployment, this setting is used  to verify the tightness of the proposed lower bound.} on the uplink WDT performance.
	Under the DSPS strategy, the uplink WDT outage probability remains unchanged as $W$ varies, consistent with \eqref{eq:SNRWPSmultipleusers}, which is independent of $W$. Under the DEPS strategy, the decrease in outage probability is minimal as $W$ increases, as the system cannot fully exploit the diversity introduced by the FA. In contrast, the outage performance under the UCPS and USPS strategies improves more noticeably as $W$ increases. This is because, when the number of ports $N$  is fixed, increasing $W$ reduces the correlation between ports (increasing $B$), resulting in better performance. However, as $W$ grows continuously, the outage probabilities under DEPS, UCPS, and USPS strategies gradually approach the derived lower bound. This occurs because, when $N$ is fixed, a larger $W$ increases the distance between ports, eliminating the correlation, making the $N$-port FA equivalent to $N$ independent single-antenna systems, as shown in Proposition \ref{EPSlowerbound}, Proposition \ref{Pro:SNRmaxG-closedform}, and Proposition \ref{SPSlowerbound}.
	
	\begin{figure}[t] 
		\centering  
		\includegraphics[width=2.2in]{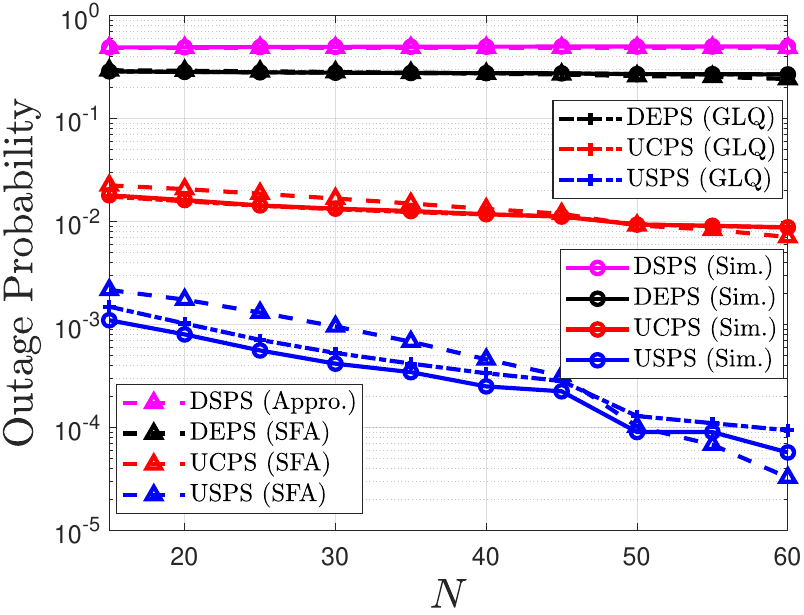}  
		\caption{Uplink WDT outage probability versus $N$; $M = 5$, $P_t=25$ dBm, $W=3$, $\gamma_{\text{th}}^{\text{U}}=8$ dB, and $d=20$.}
		\label{SNRN}  
	\end{figure} 
	
\cref{SNRN} investigates the effect of another critical parameter, the port number $N$, on the uplink WDT outage performance. Similar to \cref{SNROP-W}, under the DSPS strategy, the outage probability remains unchanged as $N$ increases, which aligns with the result in \eqref{eq:SNRWPSmultipleusers}. With a fixed FA size of $W = 3$, the value of $N$ is varied from $15$ to $60$, as the SFA-based approximation is generally accurate within this range. Under the DEPS, UCPS, and USPS strategies, the outage probability initially decreases with $N$, and then gradually saturates. This trend is observed because increasing $N$ at the beginning introduces additional diversity (\textit{i.e.,} a larger $B$). However, as $N$ increases, the ports are positioned more closely, which enhances the spatial correlation. As a result, the WDT performance becomes limited by the FA size $W$. Moreover, it is observed that the SFA-based results closely match the theoretical analysis.

	\subsection{Discussion}
	Table~\ref{tab:strategy_comparison} summarizes the characteristics of the four considered port selection strategies, offering a comparative perspective in terms of CSI requirements, signaling overhead, and their respective performances in downlink and uplink WDT.
		
		The DSPS strategy utilizes downlink CSI to select the port that maximizes the received SIR, thereby delivering the best performance in downlink WDT. However, since it does not consider the uplink channel conditions, its uplink performance is significantly limited. Alternatively, the DEPS strategy also relies on downlink CSI but aims to maximize energy harvesting power. Consequently, it offers only modest performance on both downlink and uplink links. The UCPS strategy makes decisions based solely on uplink CSI, leading to improved uplink performance compared to DEPS. However, as it does not consider downlink conditions entirely, its downlink WDT performance remains poor. Notably, UCPS strategy eliminates the need for downlink feedback, thereby reducing signaling overhead and implementation complexity. Finally, the USPS strategy jointly utilizes both downlink and uplink CSI to select the most appropriate port. While this results in the best uplink performance, the strategy suffers from poor downlink transmission and incurs the highest signaling overhead due to the dual-CSI acquisition.
	
\begin{table*}[t]
	\centering
	\caption{Comparison of Port Selection Strategies}
	\label{tab:strategy_comparison}
	\renewcommand{\arraystretch}{1.3}
	\resizebox{0.8\linewidth}{!}{%
			\begin{tabular}{ccccc}
		\toprule
			\textbf{Strategy} & \textbf{CSI Requirement} & \textbf{Signaling Overhead} & \textbf{Downlink WDT Performance} & \textbf{Uplink WDT Performance} \\
			\midrule
			\textbf{DSPS} & Downlink CSI & Low & Best & Poor \\
			\textbf{DEPS} & Downlink CSI & Low & Poor & Fair \\
			\textbf{UCPS} & Uplink CSI & Low & Poor & Suboptimal \\
			\textbf{USPS} & Downlink CSI + Uplink CSI & High & Poor & Best \\
			\bottomrule
		\end{tabular}
	}
\end{table*}

	\section{Conclusion}\label{Conclusion}
	This paper investigates the outage performance of a FAMA-assisted WPCN system under four port selection strategies: DSPS, DEPS, UCPS, and USPS, each associated with different implementation complexities and CSI requirements.   Analytical expressions for the downlink and uplink WDT outage probabilities under the four port selection strategies are derived based on the block-correlated channel model.  Simplified approximate expressions are further presented using the proposed SFA method, which significantly reduces computational complexity while maintaining accuracy. In addition, the applicable conditions for the approximations are provided to clarify their validity range. The results indicate that the DSPS strategy achieves superior downlink WDT performance, while the other three strategies deliver comparable but relatively lower performance in this regard. In contrast, the USPS strategy provides the best uplink WDT performance, with DSPS exhibiting the poorest uplink performance. Furthermore, the UCPS strategy emerges as a promising compromise, balancing uplink WDT performance and implementation complexity.
	\appendices
	\renewcommand{\theequation}{A.\arabic{equation}}
	\setcounter{equation}{0}
	\section{Proof of Lemma \ref{lemma:semiAppro}}\label{prooflemma:SFA}
	The expression $\left[1 - Q_p(a, b)\right]^L$ represents the joint cumulative distribution function (CDF) of $L$ independent and identically distributed noncentral chi-square random variables and is a monotonically decreasing function with respect to $a$. This function features a sharp  transition from 1 to 0, which motivates the use of a step function approximation. To determine the best approximation point, we aim to identify the transition point where the slope of $\left[1 - Q_p(a, b)\right]^L$ is maximized \cite{10623405}. Mathematically, we find the point at which the second derivative goes to 0 as
		\begin{equation}
			a^* = \arg\left\{\frac{\partial^2}{\partial a^2} \left[1 - Q_p(a, b)\right]^L = 0\right\}.
		\end{equation}
		
		We begin by computing the first-order derivative as
		\begin{equation}\label{eq:first_derivative}
			\frac{\partial\left[1 - Q_p(a,b)\right]^L}{\partial a}  = -L \left[1 - Q_p(a,b)\right]^{L-1} \frac{b^p}{a^{p-1}} e^{-\frac{a^2 + b^2}{2}} I_p(ab).
		\end{equation}
		For large $ab$, we apply the asymptotic approximation of modified Bessel function of the first kind $I_p(ab)$ as  \cite[Eq. (10.30.4)]{olver2010nist}
		\begin{equation}\label{AsymptoticI}
			I_p(ab) \sim \frac{e^{ab}}{\sqrt{2\pi ab}},
		\end{equation}
		which yields
		\begin{equation}
			\frac{\partial\left[1 - Q_p(a,b)\right]^L}{\partial a}  \approx -\frac{L}{\sqrt{2\pi}} \left[1 - Q_p(a,b)\right]^{L-1}  \frac{b^{p - \frac{1}{2}}}{a^{p - \frac{1}{2}}} e^{-\frac{(a - b)^2}{2}}.
		\end{equation}
		To find the approximate point, we need to compute the second-order derivative as
		\begin{equation}\label{second-order derivative}
			\begin{aligned}
				&\frac{\partial^2}{\partial a^2} \left[1 - Q_p(a, b)\right]^L \\
				&= -\frac{b^{p - \frac{1}{2}}}{a^{p - \frac{1}{2}}} e^{-\frac{(a - b)^2}{2}} \left\{ \frac{L - 1}{\sqrt{2\pi}} \left[1 - Q_p(a, b)\right]^{L - 2} \cdot \frac{b^{p - \frac{1}{2}}}{a^{p - \frac{1}{2}}} e^{-\frac{(a - b)^2}{2}} \right. \\
				&\quad \left. + \left[1 - Q_p(a, b)\right]^{L - 1} \left[ \frac{p - \frac{1}{2}}{a} + a - b \right] \right\}.
			\end{aligned}
		\end{equation}
		Then, it is reasonable to approximate $\left[1 - Q_p(a,b)\right]^{L-1} \approx \left[1 - Q_p(a,b)\right]^{L-2}$. After setting  \eqref{second-order derivative} to 0, we have
		\begin{equation}\label{linearFunction}
			\frac{L - 1}{\sqrt{2\pi}} \cdot \frac{b^{p - \frac{1}{2}}}{a^{p - \frac{1}{2}}} e^{-\frac{(a - b)^2}{2}} + \frac{p - \frac{1}{2}}{a} + a - b = 0.
		\end{equation}
		To obtain a closed-form solution, we define the following function
		\begin{equation}
			f(a) = \frac{L - 1}{\sqrt{2\pi}} \cdot \frac{b^{p - \frac{1}{2}}}{a^{p - \frac{1}{2}}} e^{-\frac{(a - b)^2}{2}} + \frac{p - \frac{1}{2}}{a} + a - b.
		\end{equation}
		By applying the first-order Taylor approximation on $f(a)$ at $a = b$, we can obtain the threshold $\delta(b, L)$ as
		\begin{equation}
			\delta(b, L) \approx b + \frac{\frac{L - 1}{\sqrt{2\pi}} b + p - \frac{1}{2}}{\frac{(L - 1)(p - \frac{1}{2})}{\sqrt{2\pi}} + \frac{p - \frac{1}{2}}{b} - b}.
		\end{equation}
		Finally, note that when $a \ll b$, the generalized Marcum-$Q$ function satisfies $Q_p(a, b) \to 0$, which implies that $[1 - Q_p(a, b)]^L \to 1$. Conversely, when $a \gg b$, we have $Q_p(a, b) \to 1$, and thus $[1 - Q_p(a, b)]^L \to 0$. Thus, the proof of Lemma \ref{lemma:semiAppro} ends.

	\renewcommand{\theequation}{B.\arabic{equation}}
	\setcounter{equation}{0}
	\section{Proof of Corollary \ref{corollary:SIRmaxSIR}}\label{prooflemma:SIRmaxSIR}
	The downlink WDT outage probability under the DSPS strategy can be reformulated as \eqref{FirstSIRmaxSIR},
	\begin{figure*}[t]
		\begin{equation}\label{FirstSIRmaxSIR}
			\begin{aligned}
				&P_{\text{DSPS}}^{\text{D}}\left(\gamma^{\text{D}}_{\text{th}}\right)\overset{\left(a\right)}{=}
				\prod_{b=1}^{B}\int_{0}^{\infty}\int_{0}^{\infty}\frac{\widetilde{r}_{b}^{M-2}e^{-\frac{\widetilde{r}_{b}+r_b}{2}}}{2^M\Gamma(M-1)}\left[1-\frac{1}{2}\int_{y=0}^{\infty}Q_{1}\left(\sqrt{\frac{\mu^{2}r_{b}}{1-\mu^{2}}},\sqrt{\gamma^{\text{D}}_{\text{th}}y}\right)\left(\frac{y\left(1-\mu^2\right)}{\mu^2\widetilde{r}_b}\right)^{\frac{M-2}{2}}\right.\\
				&\left.\quad\quad\quad\quad\quad\quad\quad\quad\quad\quad\quad\quad\quad\quad\quad\quad\quad\quad\quad\quad\times\exp\left(-\frac{y+\frac{\mu^2}{1-\mu^2}\widetilde{r}_b}{2}\right) I_{M-2}\left(\sqrt{\frac{\mu^2\widetilde{r}_by}{1-\mu^2}}\right)dy\right]^{L_b}dr_bd\widetilde{r}_{b}\\
				&\overset{\left(b\right)}{\approx}\prod_{b=1}^{B}\int_{0}^{\infty}\int_{0}^{\infty}\frac{\widetilde{r}_{b}^{M-2}e^{-\frac{\widetilde{r}_{b}+r_b}{2}}}{2^M\Gamma(M-1)}\left[1-\frac{1}{\left(\frac{\mu^2\widetilde{r}_b}{1-\mu^2}\right)^{\frac{M-2}{2}}}\int_{y=0}^{\frac{\sqrt{\frac{\mu^2r_b}{\left(1-\mu^2\right)}}+\sqrt{\frac{\mu^2r_b}{\left(1-\mu^2\right)}+2}}{2\sqrt{\gamma^{\text{D}}_{\text{th}}}}}y^{M-1}e^{-y^2-\frac{\mu^2}{1-\mu^2}\widetilde{r}_b}I_{M-2}\left(\sqrt{\frac{\mu^2\widetilde{r}_b}{1-\mu^2}}y\right)dy\right]^{L_b}dr_bd\widetilde{r}_{b}\\
				&\overset{\left(c\right)}{=}\prod_{b=1}^{B}\int_{0}^{\infty}\int_{0}^{\infty}\frac{\widetilde{r}_{b}^{M-2}e^{-\widetilde{r}_{b}-r_b}}{\Gamma(M-1)}\left[Q_{M-1}\left(\sqrt{\frac{2\mu^{2}\widetilde{r}_b}{1-\mu^{2}}},\sqrt{\frac{\mu^2r_b}{2\left(1-\mu^2\right)\gamma^{\text{D}}_{\text{th}}}}+\sqrt{\frac{\mu^2r_b}{2\left(1-\mu^2\right)\gamma^{\text{D}}_{\text{th}}}+\frac{1}{2\gamma^{\text{D}}_{\text{th}}}}\right)\right]^{L_b}dr_bd\widetilde{r}_{b}
			\end{aligned}
		\end{equation}
		\hrulefill
	\end{figure*}
	where $\left(a\right)$ uses the result of  \cite[Eq. (54)]{10623405}, $\left(b\right)$ applies the SFA and the variable substitution, and $\left(c\right)$ accounts for the fact that the generalized Marcum-$Q$-function can alternatively be defined as a finite integral. Thus, the proof of corollary \ref{corollary:SIRmaxSIR} ends.
	\renewcommand{\theequation}{C.\arabic{equation}}
	\setcounter{equation}{0}
	\section{Proof of Theorem \ref{th:SNRWPSmultipleusers}}\label{proofth:SNRmaxSIR}
	To evaluate the uplink WDT outage probability, we first exploit an important integral involving the Marcum-$Q$ function and exp established in \cite[Eq. (18)]{7105844} as
	\begin{equation}\label{Fabp}
		\begin{aligned}
			\mathcal{F}\left(a,b,c\right)&=\int_{0}^{\infty}Q_1\left(a\sqrt{x},b\right)e^{-cx}dx\\
			&=\frac{e^{-\frac{b^2}{2}}}{c}+\frac{e^{-\frac{cb^2}{a^2+2c}}}{c}\left[1-e^{-\frac{a^2b^2}{2a^2+4c}}\right].
		\end{aligned}
	\end{equation}
	Since $\mu^2$ is close to 1, we can approximate $X_n$ and $Y_n$ as follows. Specifically, $X_n$ can be approximated by $X_n\approx\left( x_{b(n)}^{(m, m, \text{D})} \right)^2 + \left( y_{b(n)}^{(m, m, \text{D})} \right)^2$, while $Y_n$ can be approximated by $
	Y_n \approx \sum_{\tilde{m} \neq m, \, \tilde{m}=1}^M \left( x_{b(n)}^{(\tilde{m}, m, \text{D})} \right)^2 + \left( y_{b(n)}^{(\tilde{m}, m, \text{D})} \right)^2$. Then, with the aid of lemma \ref{lemma:XYindependent},  $X_n + Y_n$ and $\frac{X_n}{Y_n}$ can be approximated as independent. Furthermore, due to the independence of the uplink and downlink channels, $X_n+Y_n$, $\frac{X_n}{Y_n}$ and $\beta_n$ are mutually independent. As a result, \eqref{firstSNRWPS} can be simplified as
	\begin{equation}
		P_{\text{DSPS}}^{\text{U}}\left(\gamma_{\text{th}}^{\text{U}} \right)   \approx P\left(\beta_{n^{\ast}}\alpha_{n^{\ast}}<\widetilde{\gamma}\right), \forall n^{\ast}.\\	
	\end{equation}
	After approximation, $\alpha_n=X_n+Y_n$ follows
	central Chi-square distribution, and the probability density function (PDF) of $\alpha_{n^{\ast}}$ is given by
	\begin{equation}\label{PDFalpha}
		f_{\alpha_{n^{\ast}}}(y) = \frac{1}{2^M\Gamma(M)}y^{M-1}\exp\left(-\frac{y}{2}\right), \forall n^{\ast}.
	\end{equation} 
	By defining  $\widetilde{r}=\left(x^{\left(m,\text{U}\right)}\right)^2+\left(y^{\left(m,\text{U}\right)}\right)^2$, which follows an
	exponential distribution, the unconditioned  PDF of $\beta_{n^{\ast}}$ is then expressed as
	\begin{equation}\label{PDFbetanast}
	f_{\beta_{n^{\ast}}}(t)=\int_{0}^{\infty}\frac{1}{2}e^{-\frac{\widetilde{r}}{2}}e^{-\frac{t+\frac{\mu^2}{1-\mu^2}\widetilde{r}}{2}} I_0\left(\sqrt{\frac{\mu^2\widetilde{r}t}{1-\mu^2}}\right)d\widetilde{r}, \forall n^{\ast}.
	\end{equation}
	And the unconditioned  CDF of $\beta_{n^{\ast}}$ is then formulated as
	\begin{equation}\label{CDFbeta}
		F_{\beta_{n^{\ast}}}(t)=\int_{0}^{\infty}\frac{1}{2}e^{-\frac{\widetilde{r}}{2}}
		\left[1-Q_1\left(\sqrt{\frac{\mu^2\widetilde{r}}{1-\mu^2}},\sqrt{t}\right)\right]d\widetilde{r}, \forall n^{\ast}.
	\end{equation}
	
	Then, the uplink WDT outage probability can be formulated as in \eqref{eq:SNRWPSprocess}, where $\left(a\right)$ uses the fact that $\beta_{n^{\ast}}$ and $\alpha_{n^{\ast}}$ are independent, $\left(b\right)$ uses the results of \eqref{PDFalpha} and \eqref{CDFbeta}, $\left(c\right)$  uses the fact that the total probability of a central Chi-square random variable is $1$,   $\left(d\right)$ is obtained by setting
	$a=\sqrt{\frac{\mu^2}{1-\mu^2}}$, $b=\sqrt{\frac{\widetilde{\gamma}}{y}}$ and $c=\frac{1}{2}$ in \eqref{Fabp}. Consequently, by applying the result of  \cite[Eq. (3.471.9)]{gradshteyn2014table}, the uplink WDT outage probability is obtained as in  \eqref{eq:SNRWPSmultipleusers}, which completes the proof.
	\begin{figure*}[]
		\begin{equation}\label{eq:SNRWPSprocess}
			\begin{aligned}
				P_{\text{DSPS}}^{\text{U}}\left(\gamma_{\text{th}}^{\text{U}} \right)&\overset{\left(a\right)}{\approx} P\left(\beta_{n^{\ast}}<\frac{\widetilde{\gamma}}{\alpha_{n^{\ast}}}\right)\overset{\left(b\right)}{=}\int_{\widetilde{r}=0}^{\infty}\frac{1}{2}e^{-\widetilde{r}/2}\int_{0}^{\infty}\left[1-Q_1\left(\sqrt{\frac{\mu^2\widetilde{r}}{1-\mu^2}},\sqrt{y}\right)\right]\times\frac{1}{2^M\Gamma(M)}y^{M-1}\exp\left(-\frac{y}{2}\right)dyd\widetilde{r}\\
				&\overset{\left(c\right)}{=}1-\frac{1}{2^{M+1}\Gamma(M)}\int_{\widetilde{r}=0}^{\infty}\int_{y=0}^{\infty}\exp\left(-\frac{\widetilde{r}}{2}\right)Q_1\left(\sqrt{\frac{\mu^2\widetilde{r}}{1-\mu^2}},\sqrt{\frac{\widetilde{\gamma}}{y}}\right)y^{M-1}\exp\left(-\frac{y}{2}\right)dyd\widetilde{r}\\
				&\overset{\left(d\right)}{=}1-\frac{1}{\Gamma(M)}\int_{y=0}^{\infty}y^{M-1}\exp\left(-y-\frac{\widetilde{\gamma}\left(1-\mu^2\right)}{4y}\right)dy\\
			\end{aligned}
		\end{equation}
		\hrulefill
	\end{figure*}
	\renewcommand{\theequation}{D.\arabic{equation}}
	\setcounter{equation}{0}
	\section{Proof of Theorem \ref{th:SIRmaxEHP}}\label{proofth:SIRmaxEHP}
	Similarly, since $\mu^2$ is close to 1, we can approximate $X_n$ as $	X_n\approx\left(x_{b(n)}^{(m,m,\text{D})} \right)^2+\left(y_{b(n)}^{(m,m,\text{D})} \right)^2$ and $Y_n$ as $	Y_n\approx\sum_{\tilde{m}\neq m \atop\tilde{m}=1}^M \left(x_{b(n)}^{\left(\tilde{m},m,\text{D}\right)} \right)^2+\left(y_{b(n)}^{\left(\tilde{m},m,\text{D}\right)} \right)^2$, respectively.
	Then, with the aid of Lemma \ref{lemma:XYindependent}, it can be approximated that  $X_n + Y_n$ and $\frac{X_n}{Y_n}$ are independent. Then, \eqref{eq:originalDEPS} can be simplified as
	\begin{equation}
		P_{\text{DEPS}}^{\text{D}}\left(\gamma^{\text{D}}_{\text{th}}\right)
		\approx\text{Pr}\left(\frac{X_{n^{\ast}}}{Y_{n^{\ast}}}<\gamma^{\text{D}}_{\text{th}}\right)=\text{Pr}\left(X_{n*}<Y_{n*}\gamma^{\text{D}}_{\text{th}}\right)
		,\forall  n^{\ast}.
	\end{equation} 
	Note that $X_{n^{\ast}}$ is exponentially distributed with the CDF
	\begin{equation}\label{CDF_Xnast}
		F_{X_{n^{\ast}}}\left(x\right) = 1-e^{-x/2}.
	\end{equation}
	and $Y_{n^{\ast}}$ is central Chi-squared distributed with the PDF as in \cite[Eq. (65)]{10623405} by substituting with $B=1$.
	Further, the downlink WDT outage probability by applying the DEPS strategy is then calculated as
	\begin{equation}
		\begin{aligned}
			P_{\text{DEPS}}^{\text{D}}\left(\gamma^{\text{D}}_{\text{th}}\right)&=\int_{y=0}^{\infty}F_{X_{n^{\ast}}\vert Y_{n^{\ast}}}\left(\gamma^{\text{D}}_{\text{th}} y\right)f_{Y_{n^{\ast}}}\left(y\right)dy\\
			&=1-\frac{1}{2^{M-1}\Gamma\left(M-1\right)}\int_{y=0}^{\infty}y^{M-2}e^{-\frac{y+\gamma^{\text{D}}_{\text{th}}y}{2}}dy,
		\end{aligned}
	\end{equation}
	After applying the result of \cite[eq. (3.381.4)]{gradshteyn2014table}, \eqref{result:SIRmaxEHP} is then obtained, while the proof of Theorem \ref{th:SIRmaxEHP} is completed.
	
	\renewcommand{\theequation}{E.\arabic{equation}}
	\setcounter{equation}{0}
	\section{Proof of Theorem \ref{th:SNREPSmultipleusers}}\label{proofth:SNRmaxSNR}
	Conditioned on $r_b = \sum_{\tilde{m}=1}^{M} \left(x_{b}^{\left(\tilde{m},m,\text{D}\right)}\right)^2 + \left(y_{b}^{\left(\tilde{m},m,\text{D}\right)}\right)^2$ for $ b=1,\ldots,B$, $\alpha_n$ is a non-central Chi-square variable with $2M$ degrees of freedom. The conditioned PDF of $\alpha_n$ is then obtained as 
	\begin{equation}\label{falpha}
		\begin{aligned}
			f_{\alpha_n|r_b}(x)=&\frac{1}{2}\left(\frac{x(1-\mu^{2})}{\mu^{2}r_{b}}\right)^{\frac{M-1}{2}}\exp\left(-\frac{x+\frac{\mu^{2}}{1-\mu^{2}}r_{b}}{2}\right)\\
			&\times I_{M-1}\left(\sqrt{\frac{\mu^{2}r_{b}x}{1-\mu^{2}}}\right).
		\end{aligned}
	\end{equation}
	Note that each $r_b$ for $ b=1,\ldots,B$ is a Chi-square distribution with the PDF of
	\begin{equation}
		f_{r_b}\left(r_b\right)=\frac{r_b^{M-1}e^{-r_b/2}}{2^M\Gamma\left(M\right)},b=1,\ldots,B.
	\end{equation}
	Then, the unconditioned joint CDF of $\alpha_n$ is  formulated as
	\begin{equation}\label{jointCDFalphan}
		\begin{aligned}
			&F_{\alpha_n}\left(x_1,\ldots,x_N\right)\\
			&=\prod_{b=1}^{B}\int_{0}^{\infty}\frac{r_b^{M-1}e^{-\frac{r_b}{2}}}{2^M\Gamma\left(M\right)}\prod_{n\in\mathcal{N}_b}\frac{1}{2}\left(\frac{x_n(1-\mu^{2})}{\mu^{2}r_{b}}\right)^{\frac{M-1}{2}}\\
			&\quad\quad\quad\quad\quad\quad\times\left[1-Q_{M}\left(\sqrt{\frac{\mu^{2}r_{b}}{1-\mu^{2}}},\sqrt{x_{n}}\right)\right] dr_b.
		\end{aligned}
	\end{equation}
	Thus, the uplink WDT outage probability can be reformulated as in \eqref{SNREPSprocess},
	\begin{figure*}[]
		\begin{equation}\label{SNREPSprocess}
			\begin{aligned}
				&P_{\text{DEPS}}^{\text{U}}\left(\gamma^{\text{U}}_{\text{th}}\right)=P\left(\alpha_1<\frac{\widetilde{\gamma}}{\beta_{n^{\ast}}},\ldots,\alpha_N<\frac{\widetilde{\gamma}}{\beta_{n^{\ast}}}\right)=\int_{x=0}^{\infty}F_{\alpha_n|\beta_{n^{\ast}}}(\frac{\widetilde{\gamma}}{x},\ldots,\frac{\widetilde{\gamma}}{x})\times f_{\beta_{n^{\ast}}}(x)dx\\
				&\overset{\left(a\right)}{=}\int_{\widetilde{r}=0}^{\infty}\frac{1}{2}e^{-\frac{\widetilde{r}}{2}}\int_{x=0}^{\infty}\frac{1}{2}e^{-\frac{x+\frac{\mu^{2}\widetilde{r}}{1-\mu^{2}}}{2}}I_{0}\left(\sqrt{\frac{\mu^{2}\widetilde{r}x}{1-\mu^{2}}}\right)d\widetilde{r}dx\prod_{b=1}^{B}\int_{r_b=0}^{\infty}\frac{r_b^{M-1}\exp\left(-\frac{r_b}{2}\right)}{2^{M}\Gamma(M)} \left[1-Q_M\left(\sqrt{\frac{\mu^2r_b}{1-\mu^2}},\sqrt{\frac{\widetilde{\gamma}}{x}}\right)\right]^{L_b}dr_b\\
				&\overset{\left(b\right)}{=}\int_{x=0}^{\infty}e^{-x}\int_{\widetilde{r}=0}^{\infty}e^{-\frac{1}{1-\mu^{2}}\widetilde{r}}I_{0}\left(2\sqrt{\frac{\mu^{2}\widetilde{r}x}{1-\mu^{2}}}\right)d\widetilde{r}dx\prod_{b=1}^{B}\int_{r_b=0}^{\infty}\frac{\left(r_b\right)^{M-1}\exp\left(-r_b\right)}{\Gamma(M)} \left[1-Q_M\left(\sqrt{\frac{2\mu^2r_b}{1-\mu^2}},\sqrt{\frac{\widetilde{\gamma}}{2x}}\right)\right]^{L_b}dr_b\\
			\end{aligned}
		\end{equation}
		\hrulefill
	\end{figure*}
	where $\left(a\right)$ uses the results of \eqref{CDFbeta} and \eqref{jointCDFalphan}, $\left(b\right)$ accounts for the variable substitution. Besides, due to the fact
	\begin{equation}
		2\int_{b}^{\infty}y\exp\left(-y^2-a^2\right)I_0\left(2ay\right)dy=Q_1\left(\sqrt{2}a,\sqrt{2}b\right),
	\end{equation}
	we have
	\begin{equation}\label{th:DEPSprocesssimplefied}
		\begin{aligned}		
			&\int_{\widetilde{r}=0}^{\infty}e^{-\frac{\widetilde{r}}{1-\mu^{2}}}I_{0}\left(2\sqrt{\frac{\mu^{2}\widetilde{r}x}{1-\mu^{2}}}\right)d\widetilde{r}\\
			&\overset{\left(c\right)}{=}
			\left(1-\mu^2\right)\int_{0}^{\infty}re^{-r^2}I_0\left(2\sqrt{\mu^2x}r\right)dr\\
			&\overset{\left(d\right)}{=}\left(1-\mu^2\right)\exp\left(\mu^2x\right),
		\end{aligned}
	\end{equation} 
	where $\left(c\right)$ accounts for the variable substitution, $\left(d\right)$ uses the fact that $Q_1\left(\sqrt{2\mu^2x},0\right)=1$. Finally, with the aid of \eqref{th:DEPSprocesssimplefied}, \eqref{SNREPSprocess} can be further simplified as in \eqref{eq:SNREPSmultipleusers}, which completes the proof.
	\renewcommand{\theequation}{F.\arabic{equation}}
	\setcounter{equation}{0}
	\section{Proof of Proposition \ref{EPSlowerbound}}\label{proof:EPSlowerbound}
	A simple bound for FA can be obtained by directly setting $\mu^2 = 1$, which means perfect correlation
	between the ports within each block. Then, we have
	\begin{equation}\label{boundalpha}
		\alpha_n = \sum_{\tilde{m}=1}^M \left(x_{b(n)}^{\left(\tilde{m},m,\text{D}\right)} \right)^2+\left(y_{b(n)}^{\left(\tilde{m},m,\text{D}\right)} \right)^2.
	\end{equation}
	\begin{equation}\label{boundbeta}
		\beta_n = \left(x_{b(n)}^{(m,\text{U})} \right)^2+\left(y_{b(n)}^{(m,\text{U})} \right)^2.
	\end{equation}
	Therefore, $\alpha_n$ is central Chi-square distributed with the joint CDF as
	\begin{equation}\label{CDFalphabound}
		F_{\alpha_n}\left(t_1,\ldots,t_B\right)=\prod_{n=1}^B \frac{\Phi\left(M,\frac{t_n}{2}\right)}{\Gamma\left(M\right)},
	\end{equation}
	Similarly, $\beta_{n^{\ast}}$ is exponentially distributed with the PDF as
	\begin{equation}\label{PDFbetabound}
		f_{\beta_{n^{\ast}}}\left(t\right)=\frac{1}{2}e^{-t/2}, \forall n^{\ast}.
	\end{equation}
	Then, the lower bound is calculated as
	\begin{equation}
		P_{\text{DEPS}}^{\text{U}}\left(\widehat{\gamma}\right)=\int_{0}^{\infty}F_{\alpha_{n}|\beta_{n^{\ast}}}\left(\frac{\widehat{\gamma} }{t},\ldots,\frac{\widehat{\gamma} }{t}\right) f_{\beta_{n^{\ast}}}(t) dt.
	\end{equation}
	As a result, \eqref{DEPS-lowerbound} is derived by using the results of \eqref{CDFalphabound} and \eqref{PDFbetabound}, which completes the proof.
	
	\section{Proof of Theorem \ref{th:SNRUPSmultipleusers}}\label{prooth:SNRUPSmultipleusers}
	\renewcommand{\theequation}{G.\arabic{equation}}
	\setcounter{equation}{0}
	Conditioned on $\widetilde{r}_b=\sum_{b=1}^{B}\left(x_{b}^{\left(m,\text{U}\right)}\right)^2+\left(y_{b}^{\left(m,\text{U}\right)}\right)^2$, $\beta_n$ is a noncentral Chi-square random variable with 2 degrees of
	freedom. Note that variables $\widetilde{r}_b,\forall b$ are all independent, and each follows an exponential distribution with the PDF as
	\begin{equation}
		f_{\widetilde{r}_b}\left(\widetilde{r}_b\right) = \frac{1}{2}e^{-\frac{\widetilde{r}_b}{2}}, b=1,\ldots,B.
	\end{equation}
	Then, the unconditioned joint PDF of $\beta_n$ is formulated as
		\begin{equation}\label{jointPDFbeta}
			\begin{aligned}
				&f_{\beta_n}(y_1, \ldots, y_N) = \prod_{b=1}^{B} \int_{0}^{\infty} \frac{1}{2} e^{-\widetilde{r}_b/2}
				\\&\quad\quad\times \prod_{n \in \mathcal{N}_b} \exp\left( -\frac{y_n + \frac{\mu^2}{1-\mu^2} \widetilde{r}_b}{2} \right) I_0 \left( \sqrt{\frac{\mu^2 \widetilde{r}_b y_n}{1-\mu^2}} \right) d\widetilde{r}_b.
			\end{aligned}
	\end{equation}
	And the unconditioned joint CDF of $\beta_n$ is formulated as
	\begin{equation}\label{jointCDFbeta}
			\begin{aligned}
				&F_{\beta_n}\left(y_1,\ldots,y_N\right)\\
				&=\prod_{b = 1}^{B}\int_{0}^{\infty}\frac{1}{2}e^{-\widetilde{r}_b/2}\prod_{n\in\mathcal{N}_{b}}\left[1 - Q_{1}\left(\sqrt{\frac{\mu^{2}\widetilde{r}_b}{1 - \mu^{2}}},\sqrt{y_{n}}\right)\right]d\widetilde{r}_b
			\end{aligned}
	\end{equation}
	Then, the uplink WDT outage probability in \eqref{eqfirstSNRmaxh} is derived as in \eqref{eq:SNRUPSprocess},
	\begin{figure*}[t]
		\begin{equation}\label{eq:SNRUPSprocess}
			\begin{aligned}
				P_{\text{UCPS}}^{\text{U}}\left(\widetilde{\gamma}\right) &\hspace{0.5mm}= P\left(\beta_{1}<\frac{\widetilde{\gamma}}{\alpha_{n^{\ast}}},\ldots,\beta_N<\frac{\widetilde{\gamma}}{\alpha_{n^{\ast}}}\right) =\int_{0}^{\infty}F_{\beta_n\vert\alpha_{n^{\ast}}}\left(\frac{\widetilde{\gamma}}{y},\ldots,\frac{\widetilde{\gamma}}{y}\right)f_{\alpha_{n^{\ast}}|r}(y)dy \\
				&\overset{\left(a\right)}{=} \int_{y=0}^{\infty}\int_{r=0}^{\infty} \frac{r^{M-1}e^{-\frac{r}{2}}}{2^M\Gamma(M)} \prod_{b=1}^B \int_{\widetilde{r}_b=0}^{\infty} \frac{1}{2} e^{-\frac{\widetilde{r}_b}{2}} \left[1-Q_{1}\left(\sqrt{\frac{\mu^{2}\widetilde{r}_b}{1-\mu^{2}}},\sqrt{\frac{\widetilde{\gamma}}{y}}\right)\right]^{L_b} \\
				&\quad \times \frac{1}{2} \left(\frac{y(1-\mu^{2})}{\mu^{2}r}\right)^{\frac{M-1}{2}} \exp\left(-\frac{y + \frac{\mu^{2}}{1-\mu^{2}}r}{2}\right) I_{M-1}\left(\sqrt{\frac{\mu^{2}ry}{1-\mu^{2}}}\right) dr d\widetilde{r}_b dy \\
				&\overset{\left(b\right)}{=} \left(\frac{1-\mu^{2}}{\mu^{2}}\right)^{\frac{M-1}{2}} \frac{1}{\Gamma(M)} \int_{y=0}^{\infty} \exp\left(-y\right) y^{\frac{M-1}{2}} dy \int_{r=0}^{\infty} r^{\frac{M-1}{2}} \exp\left(-\frac{r}{1-\mu^{2}}\right) I_{M-1}\left(2\sqrt{\frac{\mu^{2}ry}{1-\mu^{2}}}\right) dr \\
				&\quad \times \prod_{b=1}^B \int_{\widetilde{r}_b=0}^{\infty} e^{-\widetilde{r}_b} \left[1-Q_{1}\left(\sqrt{\frac{2\mu^{2}\widetilde{r}_b}{1-\mu^{2}}},\sqrt{\frac{\widetilde{\gamma}}{2y}}\right)\right]^{L_b} d\widetilde{r}_b
			\end{aligned}
		\end{equation}
		\hrulefill
	\end{figure*}
	where $\left(a\right)$ uses the results of \eqref{falpha} with $B=1$ and \eqref{jointCDFbeta}, $\left(b\right)$ uses the variable substitution. Then, with the aid of some simplifications, \eqref{eq:SNRUPSmultipleusers} is derived, which completes the proof.
	\section{Proof of Proposition \ref{SPSlowerbound}}\label{proof:SPSlowerbound}
	\renewcommand{\theequation}{H.\arabic{equation}}
	\setcounter{equation}{0}
	Similar to Proposition \ref{EPSlowerbound}, a bound can be obtained by directly setting $\mu^2 = 1$. Subsequently, $\alpha_{n}$ and $\beta_{n}$ can be expressed as in \eqref{boundalpha} and \eqref{boundbeta}, respectively.
	Therefore, the joint PDF of $\alpha_{n}$ is expressed as
	\begin{equation}\label{PDFboundalpha}
		f_{\alpha_n}\left(y_1,\ldots,y_B\right)=\prod_{n=1}^B \frac{1}{2^M\Gamma\left(M\right)}y_n^{M-1}e^{-y_n/2},
	\end{equation}
	and the joint CDF of $\beta_{n}$ is expressed as
	\begin{equation}\label{CDFboundbeta}
			F_{\beta_n}\left(t_1,\ldots,t_B\right)=\prod_{n=1}^B 1-e^{-t_n/2}.
	\end{equation}
	Therefore, the uplink WDT outage probability is bounded as
	\begin{multline}
		P_{\text{USPS}}^{\text{U}}(\widehat{\gamma})>\int_{0}^{\infty}\cdots\int_{0}^{\infty}F_{\beta_{n}|\alpha_{n}}(\frac{\widehat{\gamma}}{y_{1}},\ldots,\frac{\widehat{\gamma}}{ {y_{B}}})\\
		\times f_{\alpha_{n}}(y_{1},\ldots,y_{B}) dy_{1}\cdots dy_{B}.
	\end{multline}
	Finally, \eqref{USPS-lowerbound} is obtained by using the results of  \eqref{PDFboundalpha},  \eqref{CDFboundbeta}, and \cite[ Eq. (3.471.9)]{gradshteyn2014table} as well as some simplifications, which completes the proof.
	\bibliographystyle{IEEEtran}
	\bibliography{ref}
	
\end{document}